\documentclass[12pt]{article}
\usepackage{graphicx}
\usepackage{amsmath}
\def\Re{{\cal R \mskip-4mu \lower.1ex \hbox{\it e}\,}}
\def\Im{{\cal I \mskip-5mu \lower.1ex \hbox{\it m}\,}}
\def\ie{{\it i.e.}}
\def\eg{{\it e.g.}}

\def\sub#1{_{\lower.25ex\hbox{$\scriptstyle#1$}}}

\def\to{\rightarrow}

\def\subw{_{\rm w}}
\def\mh{\ifmmode m\sbl H \else $m\sbl H$\fi}
\def\mch{\ifmmode m_{H^\pm} \else $m_{H^\pm}$\fi}
\def\mt{\ifmmode m_t\else $m_t$\fi}
\def\mc{\ifmmode m_c\else $m_c$\fi}
\def\mz{\ifmmode M_Z\else $M_Z$\fi}
\def\mw{\ifmmode M_W\else $M_W$\fi}
\def\mws{\ifmmode M_W^2 \else $M_W^2$\fi}
\def\mhs{\ifmmode m_H^2 \else $m_H^2$\fi}   
\def\mzs{\ifmmode M_Z^2 \else $M_Z^2$\fi}
\def\mts{\ifmmode m_t^2 \else $m_t^2$\fi}
\def\mcs{\ifmmode m_c^2 \else $m_c^2$\fi}
\def\mchs{\ifmmode m_{H^\pm}^2 \else $m_{H^\pm}^2$\fi}
\def\ztwo{\ifmmode Z_2\else $Z_2$\fi}
\def\zone{\ifmmode Z_1\else $Z_1$\fi}
\def\mtwo{\ifmmode M_2\else $M_2$\fi}
\def\mone{\ifmmode M_1\else $M_1$\fi}
\def\tb{\ifmmode \tan\beta \else $\tan\beta$\fi}
\def\xw{\ifmmode x\subw\else $x\subw$\fi}
\def\ch{\ifmmode H^\pm \else $H^\pm$\fi}
\def\lum{\ifmmode {\cal L}\else ${\cal L}$\fi}
\def\inpb{\ifmmode {\rm pb}^{-1}\else ${\rm pb}^{-1}$\fi}
\def\infb{\ifmmode {\rm fb}^{-1}\else ${\rm fb}^{-1}$\fi}
\def\epem{\ifmmode e^+e^-\else $e^+e^-$\fi}
\def\ppb{\ifmmode \bar pp\else $\bar pp$\fi}
\def\pbp{\ifmmode ~^(\bar p^)p\else $~^(\bar p^)p$\fi}
\def\bsg{\ifmmode B\to X_s\gamma\else $B\to X_s\gamma$\fi}
\def\bsll{\ifmmode B\to X_s\ell^+\ell^-\else $B\to X_s\ell^+\ell^-$\fi}
\def\bstt{\ifmmode B\to X_s\tau^+\tau^-\else $B\to X_s\tau^+\tau^-$\fi}

\newskip\zatskip \zatskip=0pt plus0pt minus0pt
\def\matth{\mathsurround=0pt}
\def\lsim{\mathrel{\mathpalette\atversim<}}

\def\atversim#1#2{\lower0.7ex\vbox{\baselineskip\zatskip\lineskip\zatskip
  \lineskiplimit 0pt\ialign{$\matth#1\hfil##\hfil$\crcr#2\crcr\sim\crcr}}}

\renewcommand{\thefootnote}{\fnsymbol{footnote}}

\begin{document} \begin{titlepage} 
\rightline{\vbox{\halign{&#\hfil\cr
&SLAC-PUB-14477\cr
}}}
\vspace{1in} 
\begin{center}

{{\Large\bf Dissecting the Wjj Anomaly: Diagnostic Tests of a Leptophobic $Z'$}
\footnote{Work supported by the Department of 
Energy, Contract DE-AC02-76SF00515}\\}
\medskip
\medskip
\normalsize 
{\large J.L. Hewett and T.G. Rizzo{\footnote {e-mail: hewett, rizzo@slac.stanford.edu}} \\
\vskip .6cm
SLAC National Accelerator Laboratory,  \\
2575 Sand Hill Rd, Menlo Park, CA 94025, USA\\}
\vskip .5cm

\end{center} 
\vskip 0.8cm

\begin{abstract} 

We examine the scenario where a leptophobic $Z'$ boson accounts for the excess of events in the $Wjj$ channel as observed by CDF.  We assume generation
independent couplings for the $Z'$ and obtain allowed regions for the four hadronic couplings using the cross section range quoted by CDF as well as constraints
from dijet production at UA2.  These coupling regions translate into well-determined rates for the associated production of $Z/\gamma+Z'$ at the Tevatron 
and LHC, as well as $W+Z'$ at the LHC,that are directly correlated with the $Wjj$ rate observed at the  Tevatron. The $Wjj$ rate at the LHC is large and this channel 
should be observed soon once the SM backgrounds are under control. The rates for $Z/\gamma+Z'$ associated production are smaller, and these processes should 
not yet have been observed at the Tevatron given the expected SM backgrounds. In addition, we also show that valuable coupling information is obtainable from the
distributions of other kinematic variables, \eg, $M_{WZ'}$, $p_T^W$, and $\cos \theta_W^*$. Once detected, these associated production processes and the 
corresponding kinematic distributions examined here will provide further valuable information on the $Z'$ boson couplings.

\end{abstract}

\renewcommand{\thefootnote}{\arabic{footnote}} \end{titlepage}


\section{Introduction and Background}

The CDF Collaboration has reported the observation of an excess of events \cite{Aaltonen:2011mk} in the $\ell\nu jj$ channel with a statistical significance 
of 3.2$\sigma$ corresponding to 4.3 fb$^{-1}$ of integrated luminosity.  Recently, CDF has included an additional 3 fb$^{-1}$ to their data sample \cite{blois}, for a
total of 7.3 fb$^{-1}$, and the significance of this anomaly has grown to $\sim$4.8$\sigma$ ($\sim 4.1 \sigma$ including systematics).  
This is now a serious situation.   An examination of the
$m_{jj}$ distribution for these events reveals a peak that is compatible with Standard Model (SM) $WW+WZ$ production, as well as a second peak that is compatible
with a new resonance at $m_{jj} \sim 150$ GeV.  

This state of affairs has gathered much attention, even before the inclusion of the additional data sample. Skeptics have been concerned about the detailed 
shape of the Monte Carlo simulation modeling of the SM background, the jet-energy scale, as well as possible contamination from top-quark
production \cite{Sullivan:2011hu}\cite{Campbell:2011gp}.  However, the CDF Collaboration has shown \cite{blois} that neither the 
top background nor changes to the jet-energy scale is likely to account for this
excess.  Optimists have offered several new physics explanations, including a new $Z'$ boson \cite{zprime1} \cite{zprime2}, technicolor \cite{Eichten:2011sh}, 
Supersymmetry with and without R-parity conservation \cite{susy}, color octet production \cite{Wang:2011ta}, and more \cite{AguilarSaavedra:2011zy}. More recently, 
the D0 Collaboration has weighed in on this anomaly{\cite{Abazov:2011af}} and does not observe a signal at the same level as claimed by CDF in a luminosity sample 
of 4.3 fb$^{-1}$.  An understanding of this discrepancy between CDF and D0 has not yet been reached, and the situation most likely will only be clarified with results 
from the LHC. Certainly, if new physics is really present, it's cross section is most likely to be at the low end of the range discussed by CDF.

Here, we will assume the excess excess observed by CDF is due to new physics, and
we will further examine the possibility of $Z'$ production, $p\bar p \to W+Z' \to \ell\nu+jj$, as the potential source.  
Interestingly, we note that the CDF data shows a sharp dip, or valley, in the $m_{jj}$
spectrum between the first peak (\ie, SM $WW+WZ$ production) and the second peak (the hypothetical $Z'$ boson); this is the behavior that one might 
expect due to the destructive interference between the SM $W$ and $Z$ and a new gauge boson \cite{Rizzo:2006nw}.  
Clearly, this new $Z'$ boson must have very leptophobic couplings in order to evade direct production at LEPII as well as the Tevatron and LHC $Z'$ Drell-Yan 
dilepton searches.  In addition, there must be some mechanism which prohibits any significant 
$Z-Z'$ mixing ($\lsim 10^{-3}$) in order to be consistent with precision electroweak data and to avoid any `leakage' of the SM $Z$ leptonic couplings to the $Z'$.  CDF 
reports \cite{Aaltonen:2011mk} that there is no particular excess of b-quarks in the events near $m_{jj}\sim 150$ GeV, and thus we will assume that the $Z'$ decays 
democratically to all kinematically accessible hadronic states, \ie, the $Z'$ has generation-independent couplings{\footnote {Note, however, that a significant 
b-quark content for these jets, $\lsim 20-30\%$, is consistent with the existing CDF data{\cite {private,blois}}.}}. 

It has been known since long ago (in preparation for the SSC) \cite{Rizzo:1992sh}, that the associated production $W/Z/\gamma + Z'$ provides an excellent opportunity
to perform diagnostic tests on the coupling structure of a new gauge boson.  In particular, if the $Z'$ explanation for the CDF excess is correct, then one 
should at some point also observe $Z+Z'$ and $\gamma+Z'$ associated production.  As we will show below, given the CDF result, one
can make relatively definitive predictions for the rates of these processes at both the Tevatron and the LHC. 

In what follows, we will perform an analysis of the possible coupling structure and strength for the $Z'$ that is consistent with 
the data and will determine the allowed regions for the left- and right-handed $Z'$ couplings.  We will then be
armed to compute the predictions for $W/Z/\gamma+jj$ production.  We find that the rates for $Z/\gamma+jj$ are likely too small to be observed at the Tevatron
with current data samples, and that a $Z'$ in $W/Z/\gamma+jj$ could be detected at the LHC with integrated luminosities of order a few fb$^{-1}$ once SM backgrounds are 
under control.  We provide the most general expressions for these cross sections.  We also examine the $M_{WZ'}$, as well as other, kinematic distributions and
show that they can yield additional valuable coupling information, particularly for the left-handed quarks.   
Our main conclusion is that the allowed regions of the $Z'$ couplings are relatively restricted, allowing for reasonably firm predictions for the associated 
production rates and the rates for other kinematic distributions. If the CDF anomaly is due to a new $\sim 150$ GeV leptophobic $Z'$ boson, the LHC should confirm 
this signal relatively soon.

\section{Analysis}

We define the couplings of the $Z'$ to the SM quarks in a manner similar to that for the conventional SM $Z$ boson,
\begin{equation}
{\cal{L}}={{g}\over {2c_W}} \bar q \gamma^\mu \big(v'_q-a'_q \gamma_5)q~Z'_\mu ~~~(q=u,d)\,,
\end{equation}
in order to make contact with our earlier analyses \cite{Rizzo:2006nw,Rizzo:1992sh}. It will also be convenient to define the chiral coupling 
combinations $u_{L,R}=v'_u\pm a'_u$ (and similarly for $u\to d$) for the analysis below. For simplicity, 
and to avoid possible issues with Flavor Changing Neutral Currents (FCNC), 
we will assume that these couplings are generation-independent; this assumption has very little
(if any) direct impact in what follows as it is essentially only the $Z'$ couplings to the first generation quarks that determine its production cross sections at the 
Tevatron and LHC. We will, however, return to this point later below when discussing the $Z'$ total decay width.

Since the observed excess is in the proposed $W^\pm +Z'$ channel, let us first examine the differential cross section for this process; it is easily 
obtained from the expressions in the original Refs.~{\cite{Brown:1978mq},\cite{Brown:1979ux}} which describe the corresponding SM process with suitable 
simple modifications:

\begin{equation}
{{d\sigma}\over {dz}}=K_W {{G_F^2M_Z^4}\over {48\pi \hat s}} (2c_W^2)\beta_W ~\Bigg[(u_L-d_L)^2 ~X +\Big[u_L^2 ~{{\hat s^2}\over {\hat u^2}}+d_L^2 ~{{\hat s^2}
\over {\hat t^2}}\Big]~Y +2u_Ld_L~(M_W^2+M_{Z'}^2)~{{\hat s}\over {\hat u \hat t}}\Bigg] \,,
\end{equation}
where $K_W$(taken to be 1.3 in our numerical analysis) 
is a NLO K-factor, $c_W=\cos \theta_W$, $\beta_W(z)$ is the speed of the $W$ boson in the center of mass (CM) frame, $z$ is the CM scattering 
angle $\cos \theta^*$, $Y=(\hat u \hat t-M_W^2M_{Z'}^2)/\hat s^2$ and the quantity $X$ is given by the expression 
\begin{equation}
X= {{1}\over {M_W^2M_{Z'}^2}} ~\Bigg[{{1}\over {4}} (\hat u \hat t-M_W^2M_{Z'}^2) +{{1}\over {2}}(M_W^2+M_{Z'}^2)\hat s\Bigg]\,.
\end{equation}
Since the SM $W$ is purely left-handed, the right-handed couplings of the $Z'$ to the SM quarks are projected out in this amplitude so that only the 
left-handed couplings of both $u$ and $d$ appear in this expression for the cross section. It is important to note that for large values of  
$\hat s$, $X$ behaves as $\sim \hat s^2/M_W^2M_{Z'}^2 >>1$ and can provide a very significant cross section enhancement when $u_L\neq d_L$ as 
was noted numerically by some previous authors {\cite{zprime1},\cite{zprime2},\cite{Rizzo:1992sh}}.  In contrast, the other terms in the cross section are of 
order unity (or parametrically smaller) in the same limit.  As we will see below, the
presence of this term will allow for a large $W+Z'$ production rate, without necessarily
enhancing the corresponding $Z/\gamma+Z'$ cross sections.
However, we note that 
the possibility of $u_L\neq d_L$ implies that the group generator, $Q'$, to which the $Z'$ couples does not commute with the usual $SU(2)_L$ isospin generators, 
\ie, $[Q',T_i]\neq 0$.  This can have a number of implications elsewhere \cite{Hewett:1992nf}. 

\begin{figure}[htbp]
\centerline{
\includegraphics[width=8.0cm,angle=90]{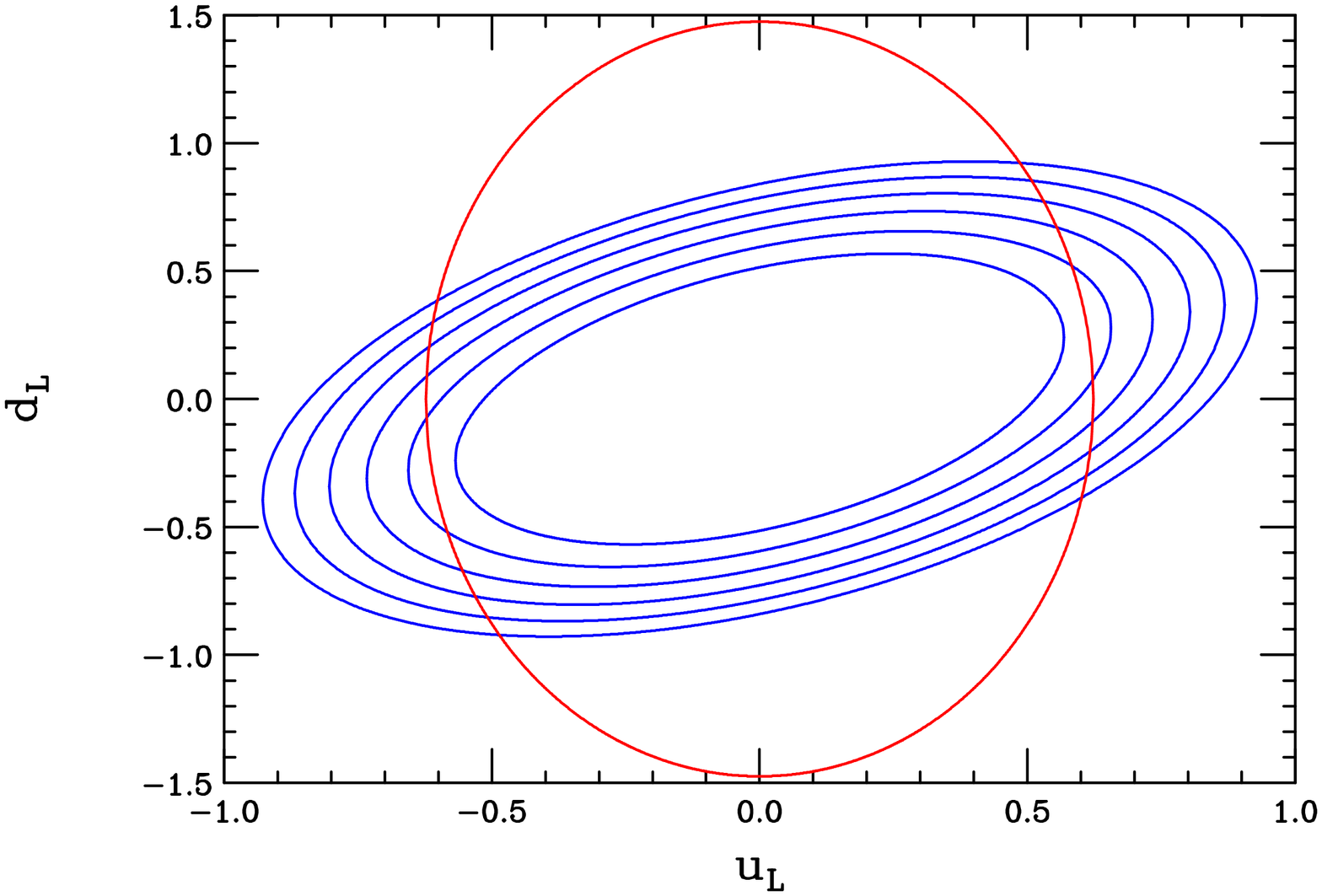}}
\vspace*{0.5cm}
\centerline{
\includegraphics[width=8.0cm,angle=90]{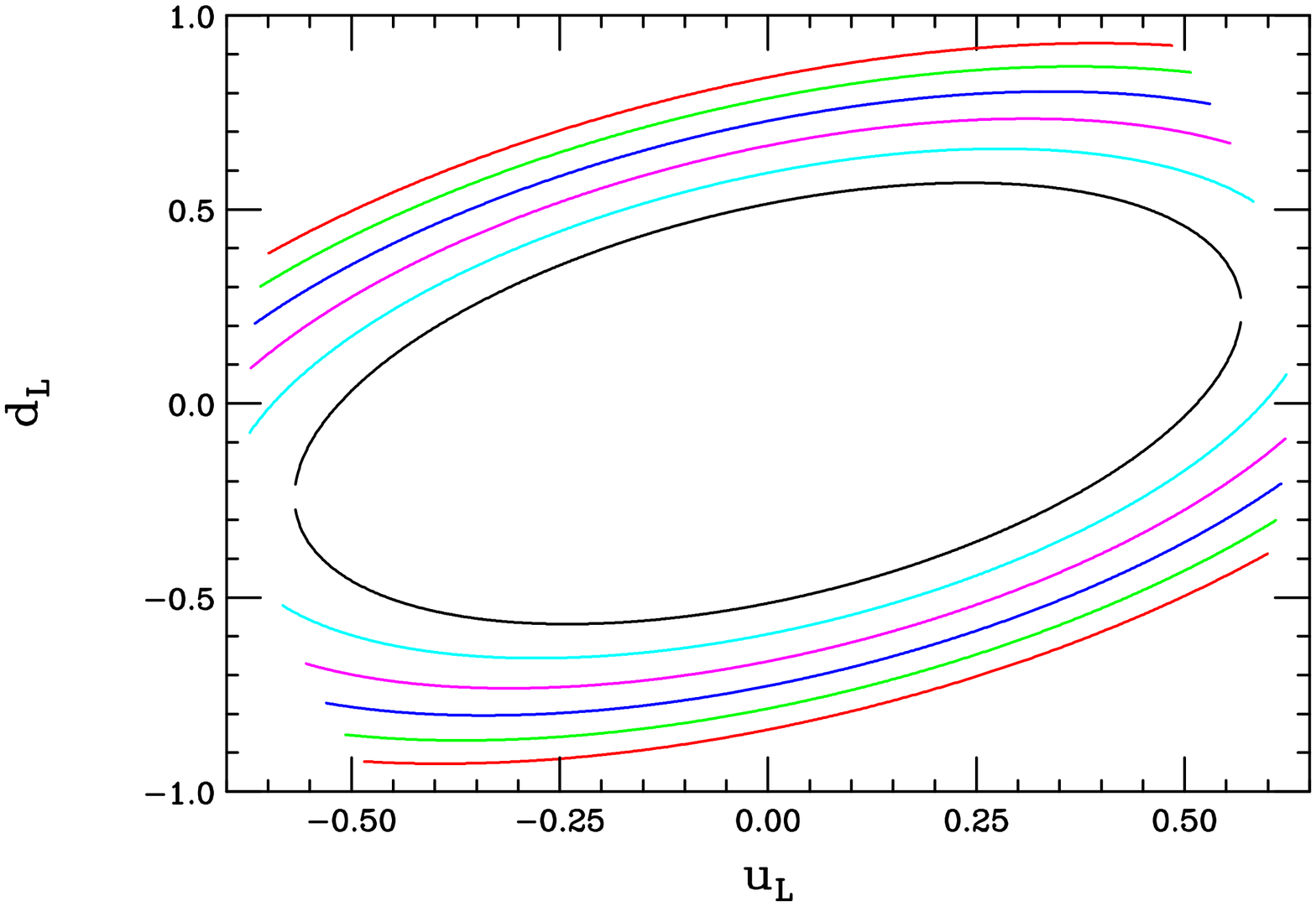}}
\vspace*{0.5cm}
\caption{Top: The blue ellipses show the values of the $u_L,d_L$ couplings leading to  the Tevatron $W^\pm Z'$ cross section of (from inside out) 1.5-4.0 
pb in steps of 0.5 pb. The red ellipse shows the {\it maximum} size of the UA2 allowed region in the $u_L-d_L$ plane. Bottom: The color-coded arcs shown here 
are the line segments where the Tevatron (blue) ellipses intersect and are contained within the UA2 (red) ellipse in the top panel.}
\label{ellipses}
\end{figure}

Requiring that 
the $Z'$ decays only to two jets, integration of the above expression over $z=\cos \theta^*$ and the relevant parton densities leads to the numerical value 
for the (pre-cut) $W+Z'$ cross sections at the Tevatron and LHC for arbitrary couplings given by 
\begin{eqnarray}
\sigma_{W^\pm Z'} & \simeq & 4.945~(u_L-d_L)^2+0.719~(u_L^2+d_L^2)+5.083~u_Ld_L ~~(\rm {pb)\quad (Tevatron)}\,,
\nonumber\\
\sigma_{W^\pm Z'} & \simeq & 28.61~(u_L-d_L)^2+4.029~(u_L^2+d_L^2)+21.65~u_Ld_L ~~(\rm {pb)\quad (LHC)}\,.
\end{eqnarray}
These results explicitly show the enhancement arising in the case of $u_L\neq d_L$. 
In performing these numerical calculations, and the ones found below, we make use of the CTEQ6.6M parton density functions {\cite{Nadolsky:2008zw}}. 
Since the apparent excess in the Tevatron 
$W+Z'$ cross section is observed {\cite{Aaltonen:2011mk}}, prior to acceptance and analysis cuts, to be in the range of $\sim 1-4$~pb, this results in an ellipse 
of potentially allowed values in the $u_L-d_L$ plane{\footnote {The exact result is somewhat sensitive to acceptance corrections.}}.  This is displayed in the 
top panel of Fig.~\ref{ellipses}, assuming $M_{Z'}=150$ GeV; in this figure we show the allowed ellipses for $W+Z'$ cross section values ranging from 1.5-4.0 pb.
Of course, given the results from D0, the lower end of this range will be likely to be of interest to us in what follows.

Of course, a leptophobic $Z'$ boson will also be produced directly and contribute to dijet production and may be
observed as a resonance in the dijet invariant mass spectrum.  Due to kinematics, the data from $Sp\bar pS$ has the best signal to background ratio for searches in 
the dijet 
channel in this low mass region.  UA2 {\cite {UA2}} performed such a search in the dijet channel and constrained the cross section to be less than 
roughly $\simeq 150$ pb for a $\sim 150$ GeV resonance. This places an additional constraint on the $Z'$ couplings that needs to be satisfied. 
Employing the narrow width approximation (which we will justify below), the dijet rate induced by a $Z'$ at UA2 (recall the CM energy for the $Sp\bar pS$ 
was 630 GeV) resulting from the process $q\bar q\to Z' \to jj$ can be written numerically as 
\begin{equation}
\sigma_{UA2} \simeq {{1}\over {2}}~[773~(u_L^2+u_R^2)+138~(d_L^2+d_R^2)] ~~(\rm {pb})\,,
\end{equation}
making use of the same procedure and assumptions as above. Given this result and the UA2 bound on the cross section, the {\it largest} corresponding constraint 
ellipse that can be drawn in the $u_L-d_L$ plane denoting the UA2 allowed region is obviously obtained when the $Z'$ has only left-handed quark couplings, 
\ie, $u_R=d_R=0$.  This bound is shown as the red ellipse in the top panel of Fig.~\ref{ellipses}. Clearly, if
non-zero values of $u_R$ or $d_R$ are also present, then this constraint ellipse will only {\it contract}.  Here we note that the 
UA2-allowed coupling ellipse intersects the corresponding ones obtained by evaluating the $W^\pm Z'$ cross section at the Tevatron at different values of the $Z'$ 
couplings depending upon the assumed value of 
$\sigma_{W^\pm Z'}$. Note that the simultaneous consistency of the CDF result with the UA2 dijet data forbids very large $u_L$ couplings of either sign and
allows for the possibility of $u_L\neq d_L$.  
The segments of these ellipses that are simultaneously allowed by both cross section constraints are highlighted in the lower panel 
of Fig.~\ref{ellipses} and are color coded for comparisons with results to be shown in later figures. 

\begin{figure}[htbp]
\centerline{
\includegraphics[width=8.0cm,angle=90]{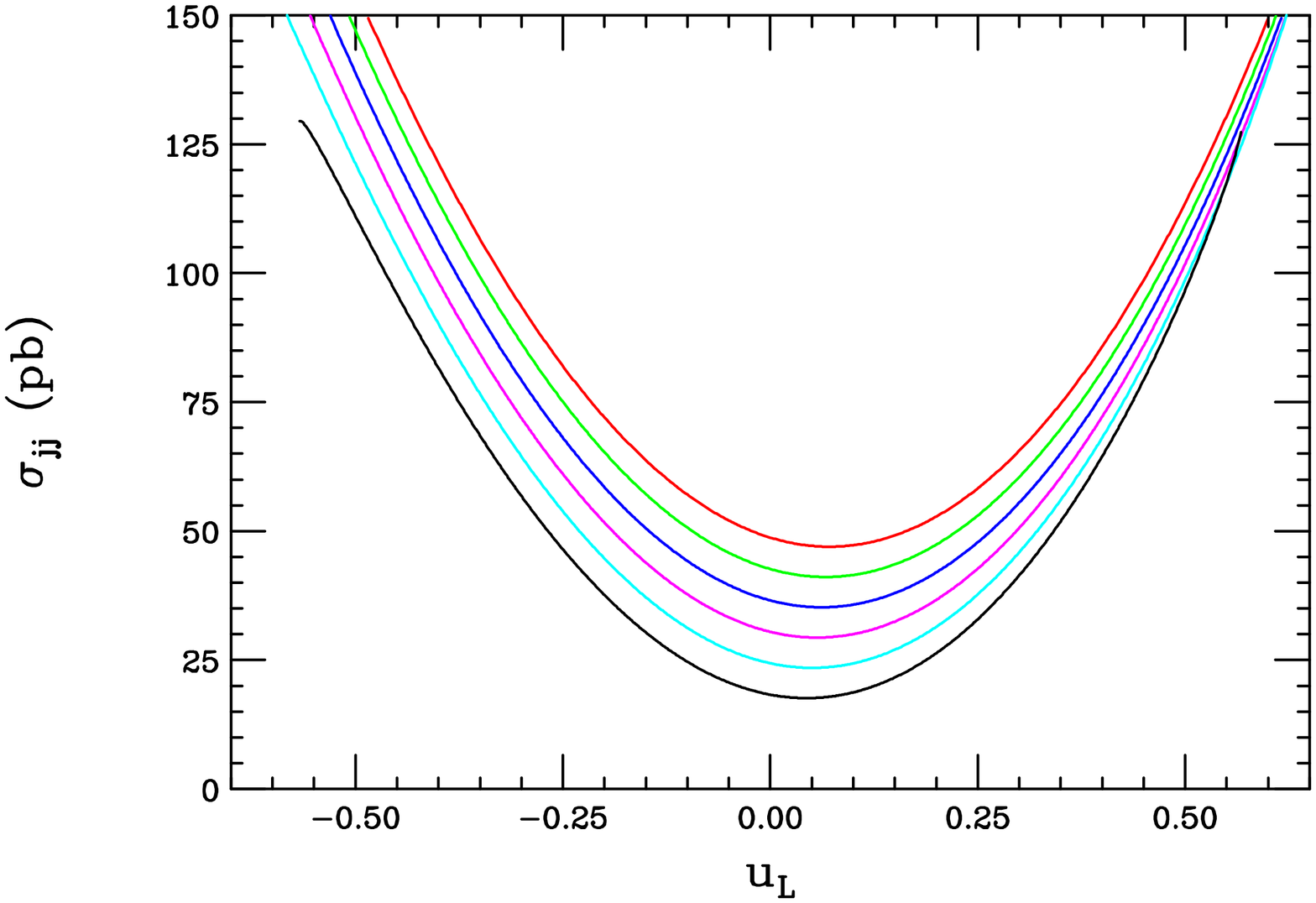}}
\vspace*{0.5cm}
\centerline{
\includegraphics[width=8.0cm,angle=90]{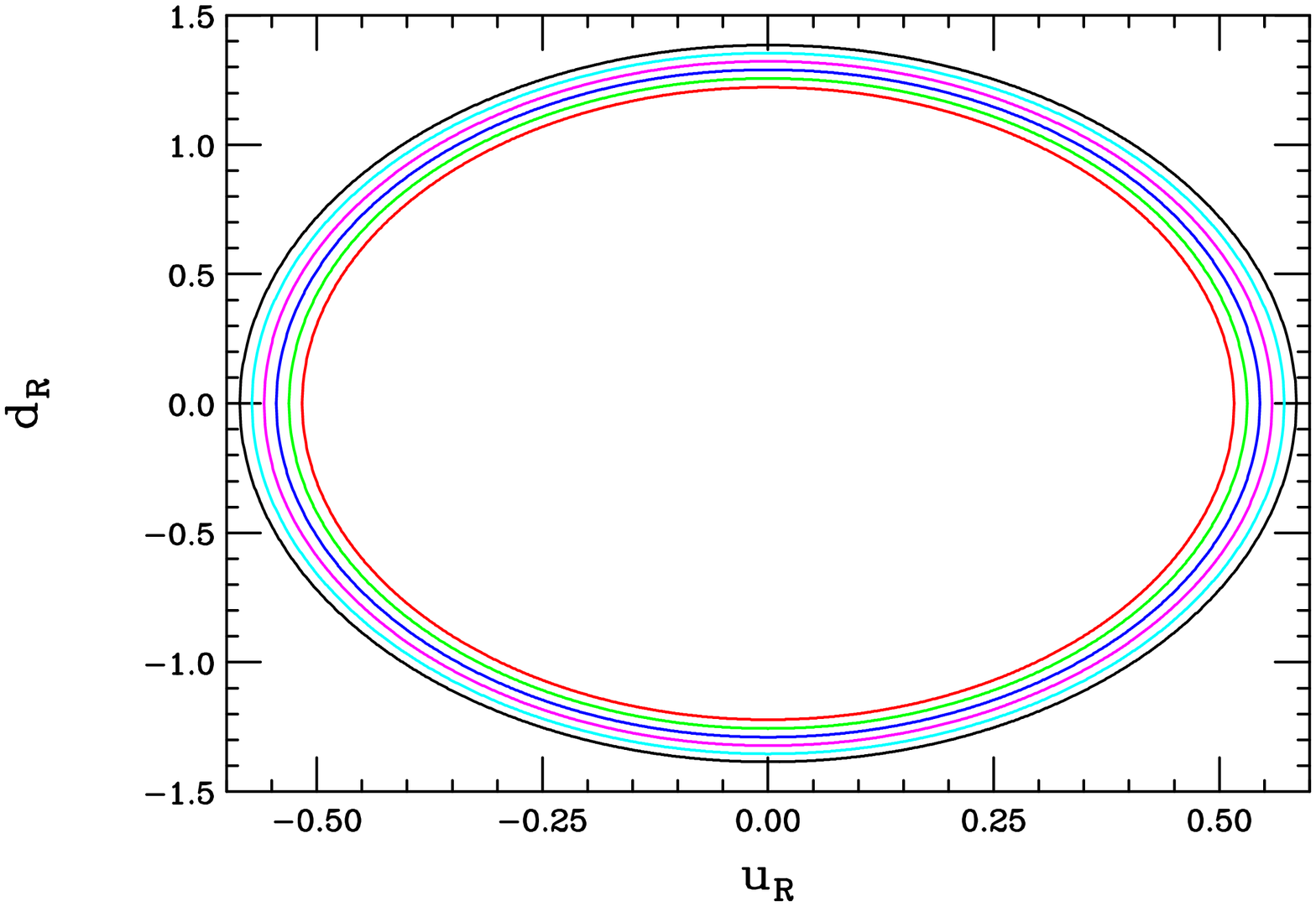}}
\vspace*{0.5cm}
\caption{Top: Predicted values of the UA2 dijet cross section along the color-coded arcs for
$u_L\,, d_L$ as shown in the previous Figure. Note that an additional set of solutions are present when $u_L \to -u_L$ as discussed in the text.
Bottom: Maximum allowed values of $u_R-d_R$ at the minima of the parabolas in the top panel. From inside out these correspond to Tevatron $W+Z'$ cross sections 
of 1.5, 2., 2.5, 3, 3.5 and 4 pb, respectively.}
\label{ua2}
\end{figure}

For the case $u_R=d_R=0$, the upper panel in Fig.~\ref{ua2} shows the predicted UA2 dijet cross section along the allowed coupling line segments of 
Fig.~\ref{ellipses}.  The curves in this figure correspond to the upper set of arcs in Fig.~\ref{ellipses}; a corresponding set of curves can also
be obtained representing the bottom arcs and is obtained by flipping the values $u_L\to -u_L$ in Fig.~\ref{ua2}.
In all cases we see that the values along the parabolic shaped curves can lead to a dijet cross section that is 
significantly far from the quoted upper 
bound. However, this still implies that the possible values of $u_R,d_R$ must be restricted or the UA2 dijet bound would be exceeded.
Of course, for any arbitrary point along these parabolas one can perform a scan of the $u_R-d_R$ plane to obtain the 
corresponding region which is allowed by UA2; the weakest bounds on the 
right-handed couplings are clearly obtained when the predicted dijet cross section is minimized. These constraints on the maximal values
of the right-handed couplings are shown in the
$u_R-d_R$ plane in
the lower panel of Fig.~\ref{ua2} for various values of the Tevatron $W+Z'$ cross section. 
Note that while the largest $u_R-d_R$ allowed region is obtained at the minimum of the parabolas in the top panel, the region will 
shrink substantially at points where the dijet cross section arising from the left-handed couplings alone almost saturates the UA2 bound. 

Now that we have obtained constraints on the left- and right-handed $Z'$ couplings,
let us turn to the other relevant processes for associated production, namely $Z+Z'$ and $\gamma +Z'$. In analogy with our $W^\pm +Z'$ 
result above, the $q\bar q \to Z+Z'$ differential cross section can be obtained by a suitable modification of the corresponding result in the SM given 
by {\cite{Brown:1978mq} \cite{Brown:1979ux}}
\begin{equation}
{{d\sigma}\over {dz}}=K_Z {{G_F^2M_Z^4}\over {48\pi \hat s}} \beta_Z ~\Big[(v_q+a_q)^2q_L^2+(v_q-a_q)^2q_R^2\Big]~P \,,
\end{equation}
where $(v,a)_q$ are the couplings of the quarks to the SM $Z$ boson, $K_Z, \beta_Z$ are the K-factor (=1.3 here) 
for this process and speed of the SM $Z$ in the CM frame, 
and $P$ represents the same kinematics as in the $W^\pm + Z'$ case above in the limit of equal couplings and with the replacement $M_W\to M_Z$, \ie , 
\begin{equation}
P=\Big(\hat u \hat t-M_Z^2M_{Z'}^2\Big)  \Big({{1}\over {\hat u^2}}+{{1}\over {\hat t^2}}\Big)+2 {{\hat s (M_Z^2+ M_{Z'}^2)}\over {\hat u \hat t}}\,.
\end{equation}
Since the SM $Z$ couplings are known, and folding in the SM $Z$ decay to lepton pairs (with $B=0.03366$ for $e$ or $\mu$ and then summing over both) this expression 
can be 
numerically evaluated for arbitrary $Z'$ couplings after integration over $z$ and the relevant parton densities at either the Tevatron or the LHC. Writing 
\begin{equation}
\sigma_{ZZ'}\simeq {{1}\over {4}} \Big[ \alpha_Z u_L^2 +\beta_Z u_R^2 +\gamma_Z d_L^2 +\delta_Z d_R^2\Big]\,,
\end{equation}
we obtain, in units of fb and before any cuts, $\alpha_Z=381.5(1109)$, $\beta_Z=221(76.1)$, $\gamma_Z=1323(166.4)$ and $\delta_Z=44.1(5.54)$ for the case of the 
Tevatron(LHC). It is important to note that the SM $Z$ leptonic branching fractions have been included here to ease comparison with experiment. 

Analogously, we can obtain the corresponding numerical result for the case of the $\gamma +Z'$ final state; the analytic expression for the 
differential cross section can be obtained from that for $Z+Z'$ production by taking $M_Z\to 0$ in $P$ and by a setting $v_Q \sim Q_q$ with $a_q=0$. 
In this case we impose the experimental cuts $|\eta_\gamma|<1.1(2.5)$ and $p_T^\gamma >25(50)$ GeV at the Tevatron(LHC) and obtain numerically after integration
\begin{equation}
\sigma_{\gamma Z'}\simeq {{1}\over {2}} \Big[ f_u^\gamma (u_L^2+u_R^2) +f_d^\gamma (d_L^2+d_R^2)\Big]\,,
\end{equation}
where $f_u^\gamma=767(533)$ fb and $f_d^\gamma=72.7(114)$ fb at the Tevatron(LHC). 

\begin{figure}[htbp]
\centerline{
\includegraphics[width=8.0cm,angle=90]{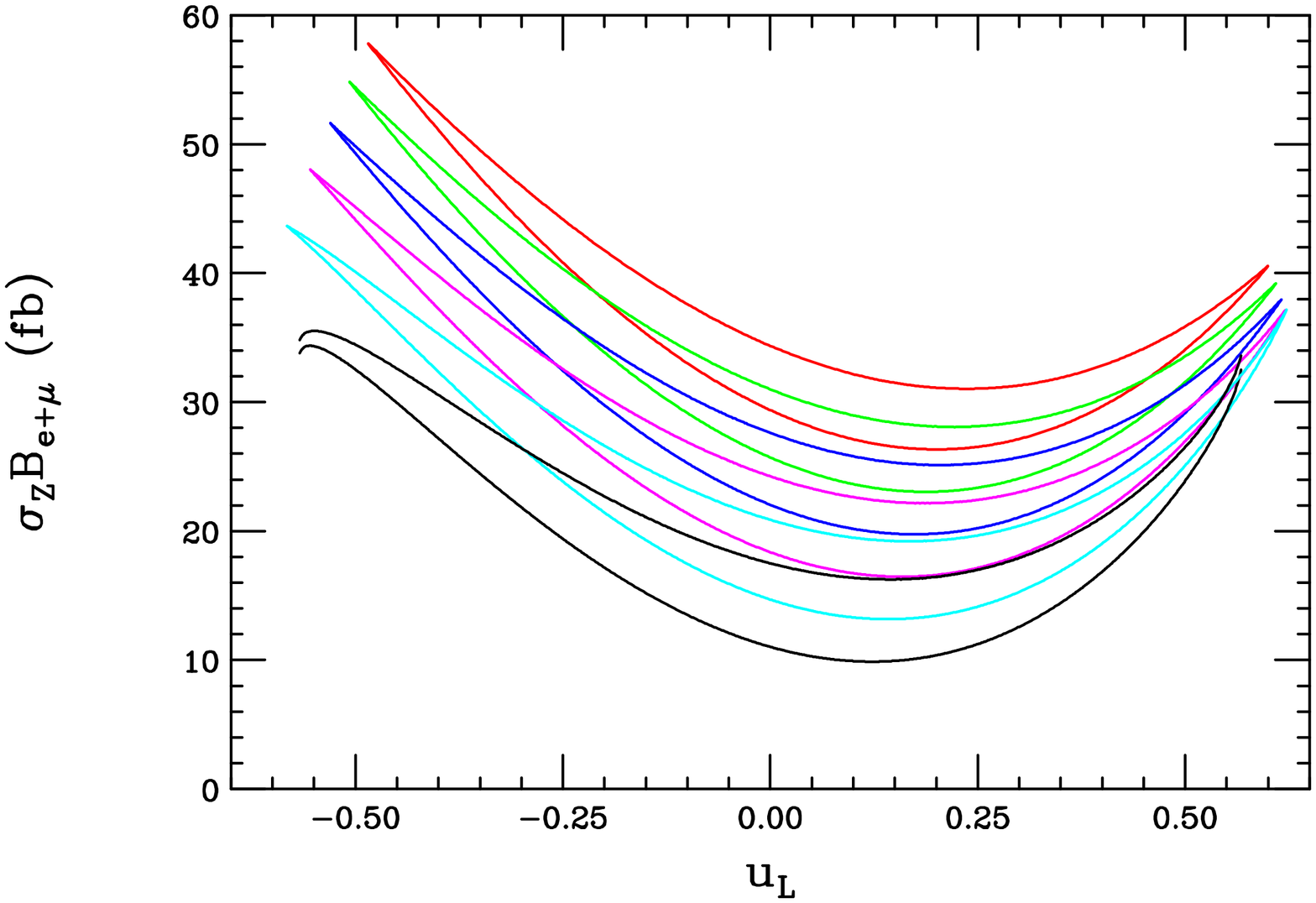}}
\vspace*{0.5cm}
\centerline{
\includegraphics[width=8.0cm,angle=90]{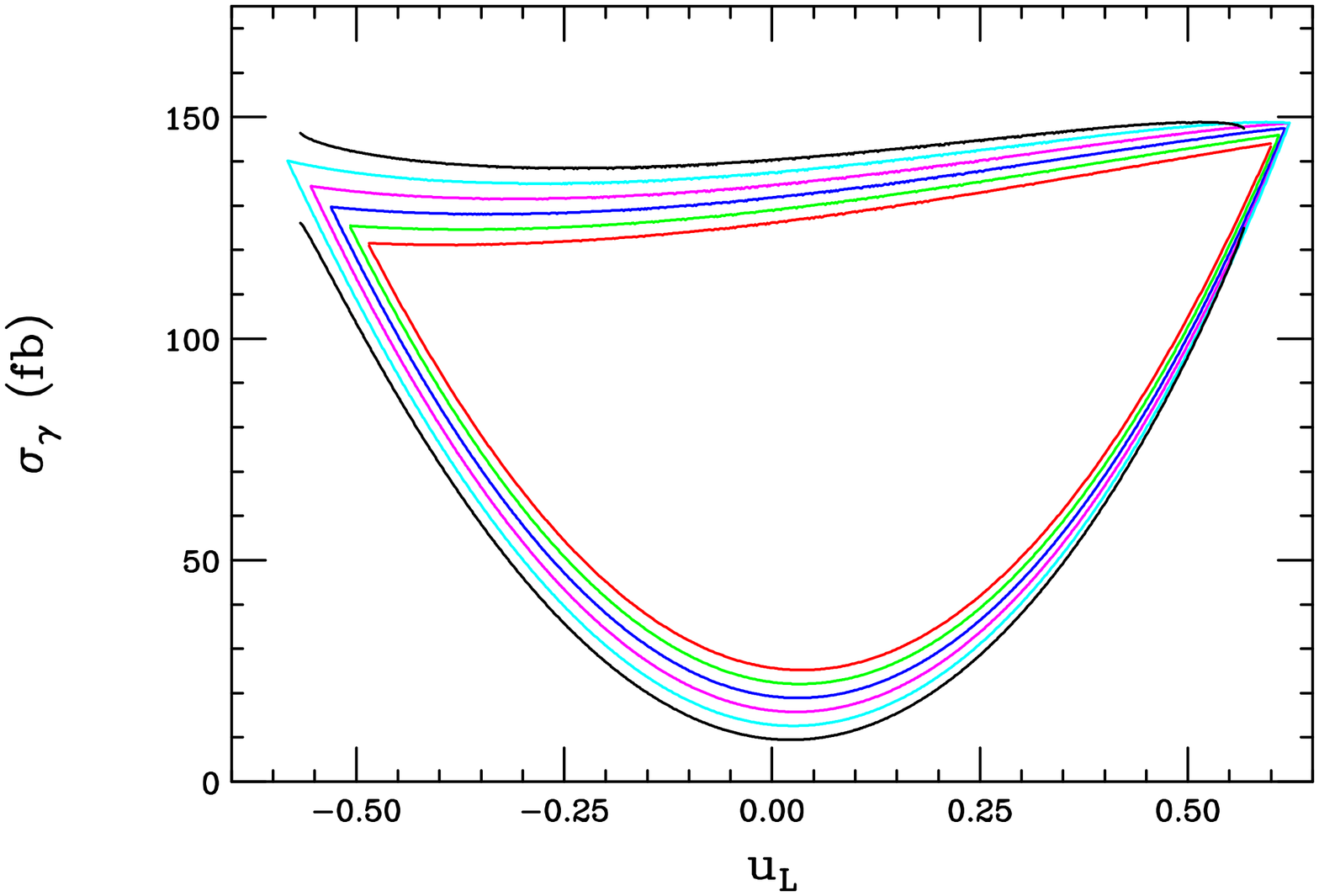}}
\vspace*{0.5cm}
\caption{Predicted allowed values of the cross sections $\sigma_{ZZ',\gamma Z'}$ at the Tevatron for the parameter space regions associated with the color-coded 
arcs shown in the previous Figures corresponding to $WZ'$ cross section of $1.5-4.0$ pb at the Tevatron.
The allowed region lies within the areas outlined in a specific color. For the $ZZ'$ final state the branching fractions for 
leptonic decay of the $Z$ are included. Note that an additional region exists with $u_L\to -u_L$ as discussed above.}
\label{tev}
\end{figure}

\begin{figure}[htbp]
\centerline{
\includegraphics[width=8.0cm,angle=90]{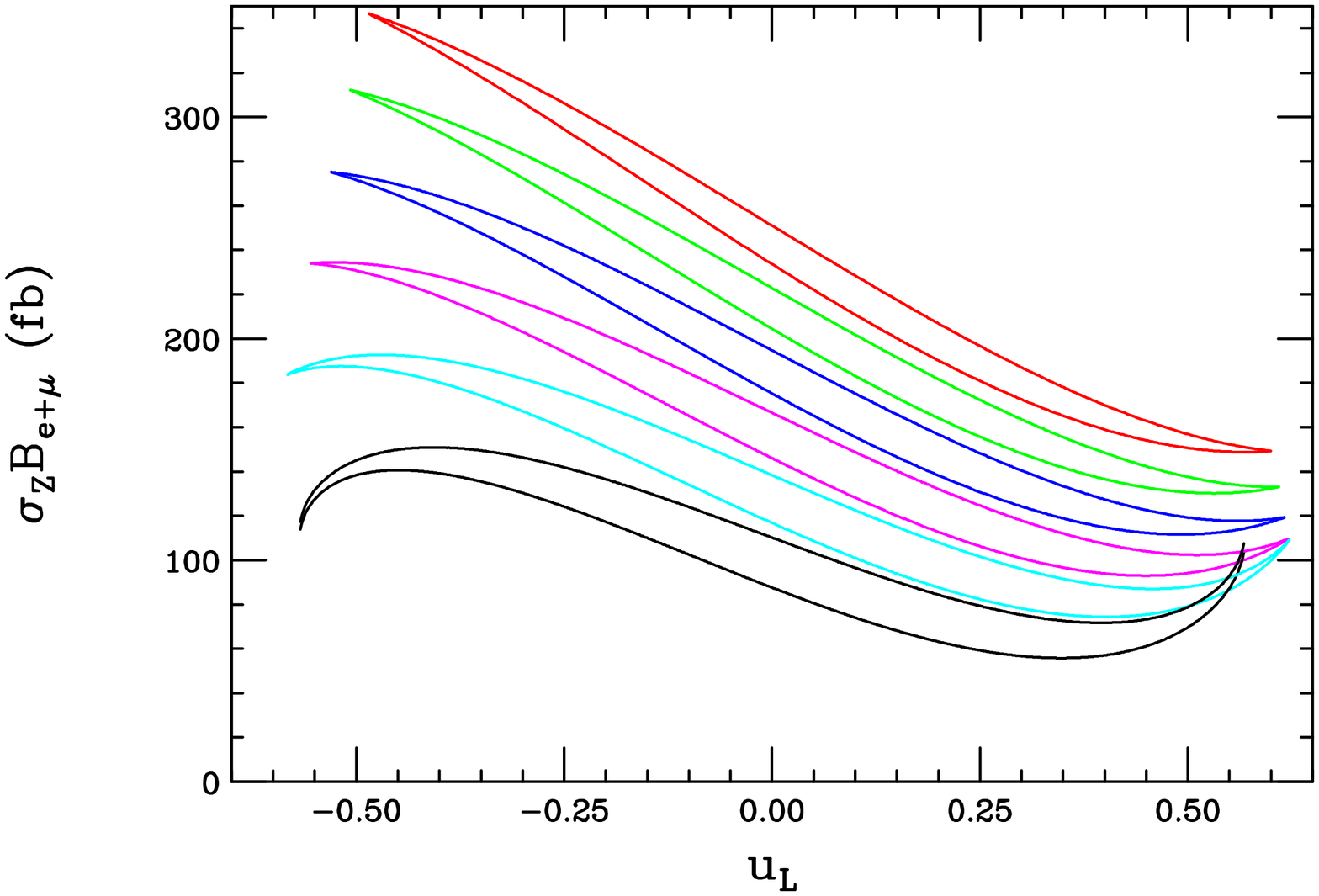}}
\vspace*{0.5cm}
\centerline{
\includegraphics[width=8.0cm,angle=90]{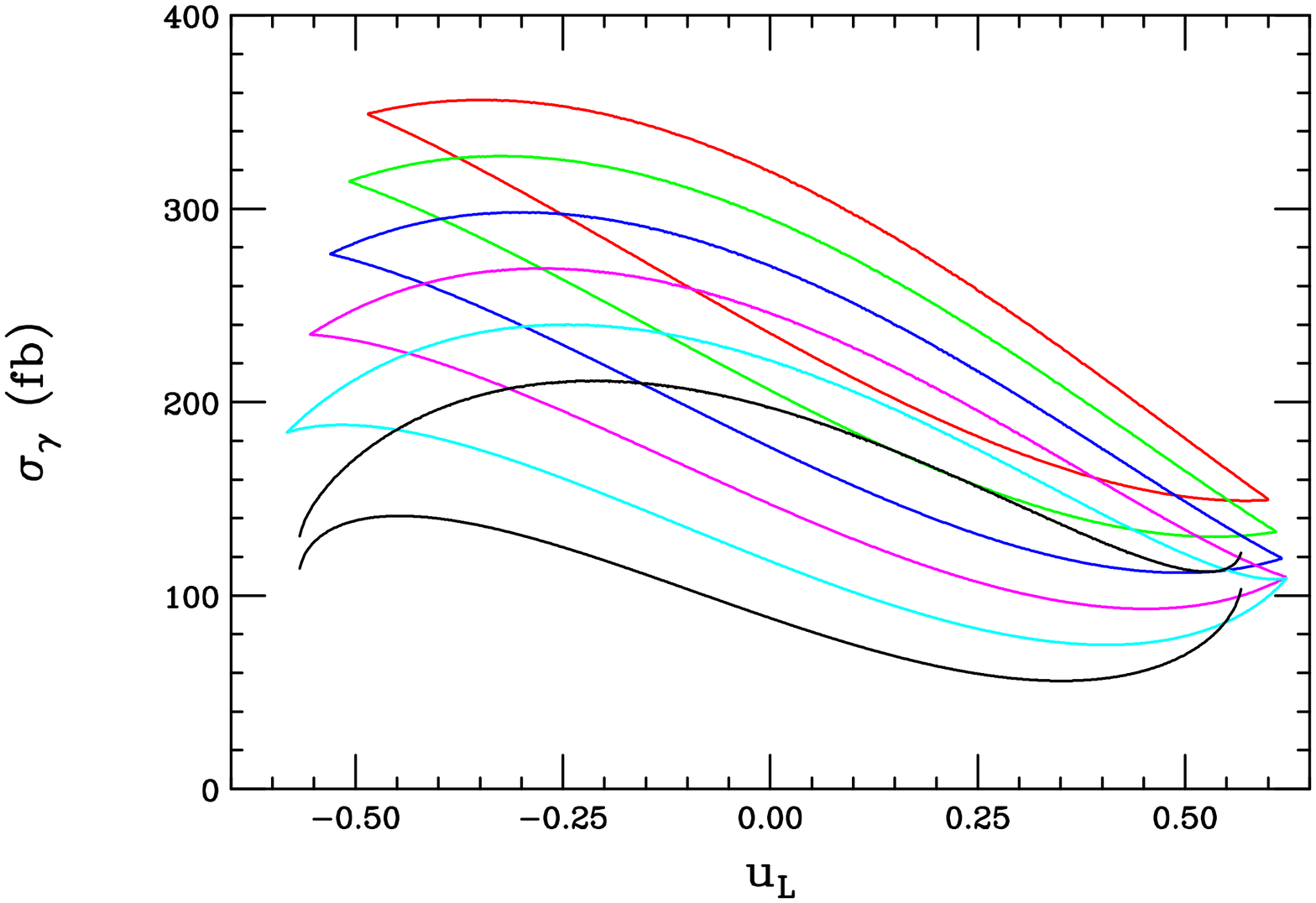}}
\vspace*{0.5cm}
\caption{Same as the previous Figure but now for the LHC.}
\label{LHC}
\end{figure}

\begin{figure}[htbp]
\centerline{
\includegraphics[width=8.0cm,angle=90]{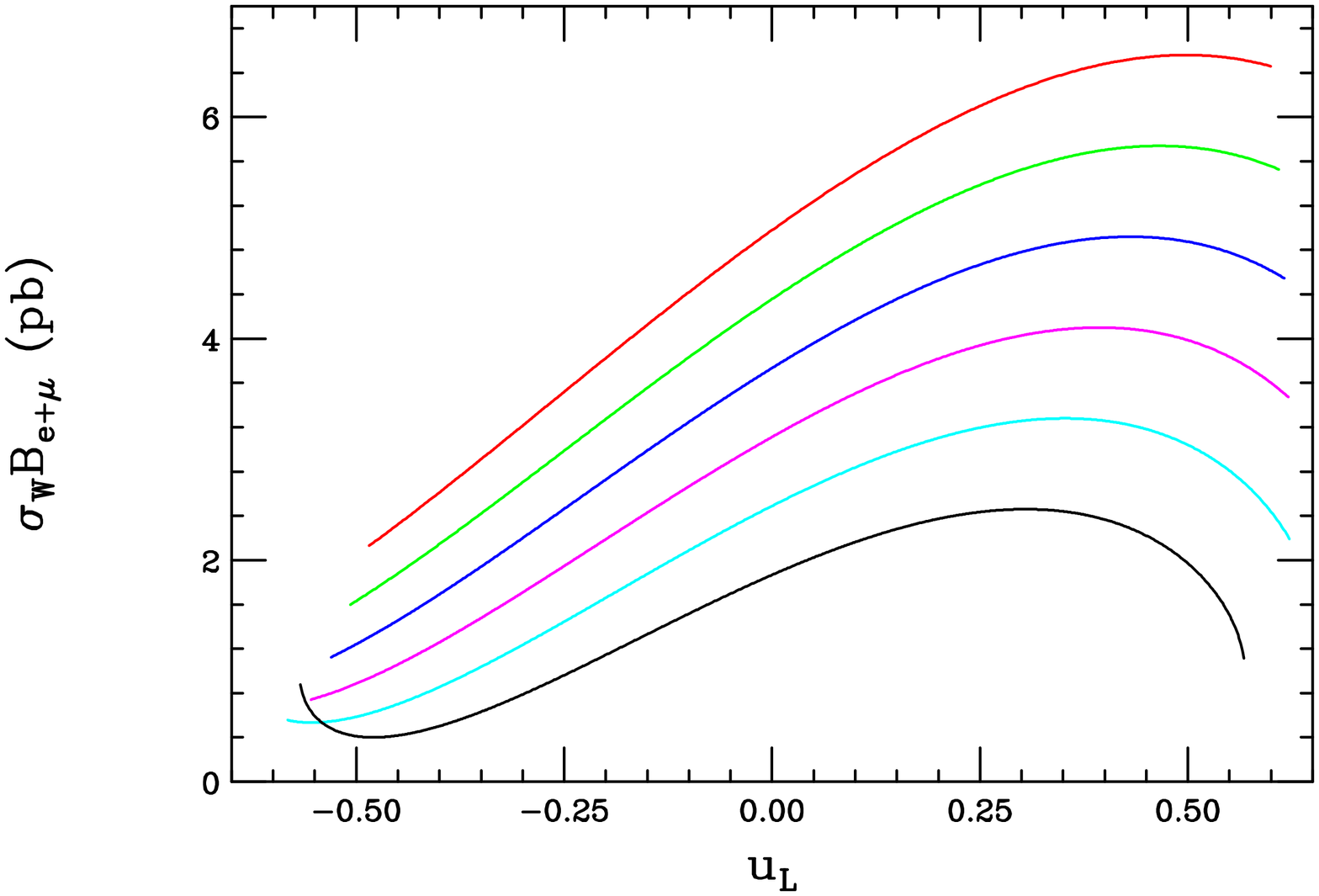}}
\vspace*{0.5cm}
\centerline{
\includegraphics[width=8.0cm,angle=90]{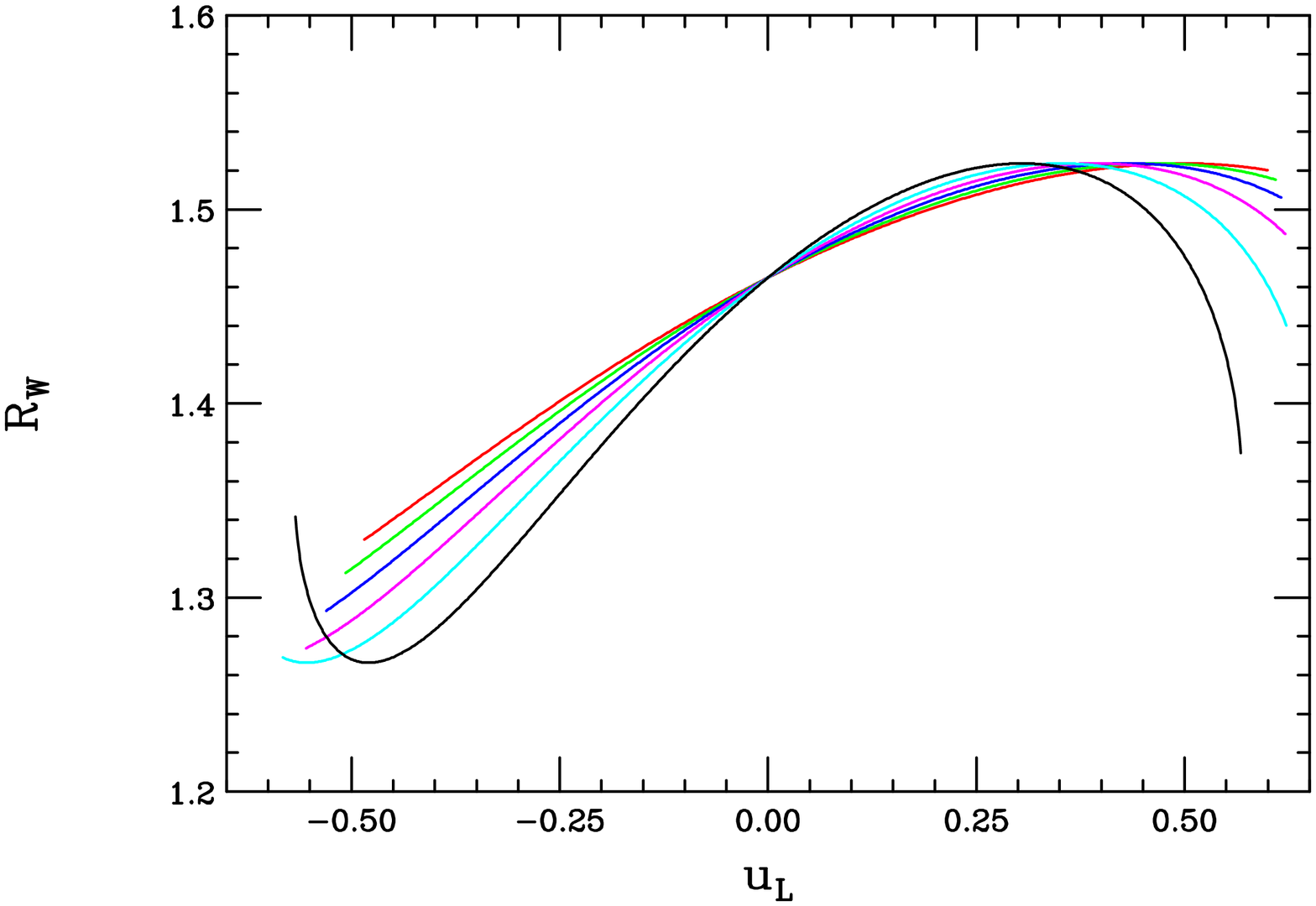}}
\vspace*{0.5cm}
\caption{(Top) The predicted values for the sum of the cross section for $W^\pm Z'$ at the LHC based on the corresponding cross section observed by CDF at the Tevatron 
along the parameter space arcs described above. The $W$ leptonic branching fraction is included.
Again, note that another set of solutions exist with $u_L\to -u_L$. (Bottom) The ratio of the $W^+Z'$ to the $W^-Z'$ cross sections at the LHC, with the same color
coding. }
\label{wzp}
\end{figure}

We are now ready to calculate the expected values of $\sigma_{ZZ',\gamma Z'}$ at both colliders. In evaluating the $Z'$ couplings, we proceed as follows: we select 
a point on one of the 
line segments in the bottom panel of Fig.~\ref{ellipses} which tells us the specific values of $u_L,d_L$. We then locate that point on the upper panel in 
Fig.~\ref{ua2} and scan over the possible values in the $u_R-d_R$ plane which are consistent with the UA2 upper bound on the dijet cross section for those 
$u_L,d_L$ couplings 
and obtain the maximum and minimum values for both $\sigma_{ZZ',\gamma Z'}$ at the Tevatron and the LHC. The minimum values in all cases correspond, of course,  
to the situation when $u_R=d_R=0$ as contributions arising from non-zero values of these couplings always add constructively. The results of this analysis for 
the Tevatron and LHC are shown in Fig.~\ref{tev} and Fig.~\ref{LHC}, respectively,
employing the same color coding as before.  We see that these cross sections are much
smaller than that for $W^\pm+Z'$ at the Tevatron (as well as for the LHC) and are possibly too
small to be observed at the Tevatron with present integrated luminosities given SM backgrounds.  These
cross sections are, of course, much larger at the LHC and should be observable with
roughly 1 fb$^{-1}$ of integrated luminosity once SM backgrounds are sufficiently understood. The predicted results 
for $\sigma_{W^\pm Z'}$ at the LHC, which are independent 
of the possible values of $u_R,d_R$ as was the case for the Tevatron, can be found in Fig.~\ref{wzp}. Note that the branching fraction for the leptonic decays of 
the SM $W$ are included in these results. We see that the cross section is quite large and should be
detectable soon.

We learn a number of things from examining these Figures: ($i$) The predicted values for $\sigma_{\gamma Z',ZZ'}$ at the Tevatron (and the LHC) are always 
substantially lower than the corresponding ones for $\sigma_{W^\pm Z'}$. These processes should {\it not yet} have been observed at the Tevatron but will 
eventually provide a test of the $Z'$ hypothesis once enough data accumulates at
the LHC. {\footnote {In fact, their observation at the Tevatron at relatively low luminosity would likely have ruled out the $Z'$ hypothesis.}} 
($ii$) The predicted values of $\sigma_{\gamma Z',ZZ'}$ are relatively constrained and are determined by the CDF $W+Z'$ cross section itself, as well as by 
the UA2 dijet constraints, except for possible NLO contributions. ($iii$) The $\gamma Z',ZZ'$ cross 
sections at the LHC and the $ZZ'$ cross section at the Tevatron are found to be relatively insensitive to the specific values of $u_R,d_R$ due to the rather 
strong constraints arising from the UA2 data. ($iv$) The $\gamma Z'$ process at the Tevatron could potentially be used to obtain further constraints on the values of 
$u_R,d_R$ given sufficient integrated luminosity. ($v$) The $W^\pm Z'$ cross 
section at the LHC is large and is well-predicted apart from possible NLO contributions. Lastly, ($vi$) we see that the {\it ratio} of the $W^+Z'$ and $W^-Z'$ 
cross sections at the LHC also has a weak coupling dependence which may also be useful as an additional handle on the left-handed quark couplings of the $Z'$. 
Thus we see that even with four free coupling parameters, the $Z'$ 
explanation of the $Wjj$ excess seen by CDF leads to a very predictive scenario that can be further tested quite soon at both the Tevatron and the LHC.


\begin{figure}[htbp]
\centerline{
\includegraphics[width=8.0cm,angle=90]{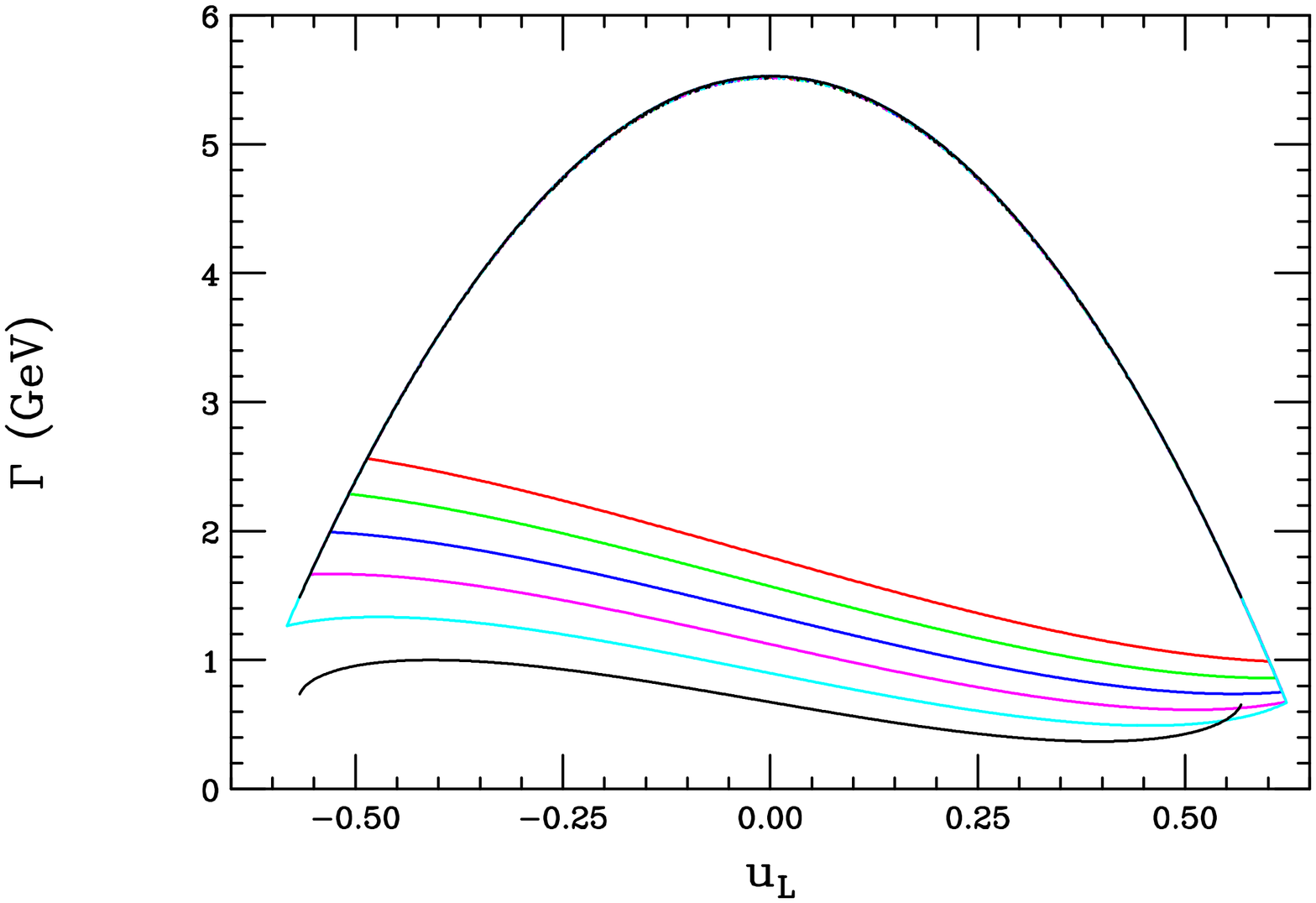}}
\vspace*{0.5cm}
\centerline{
\includegraphics[width=8.0cm,angle=90]{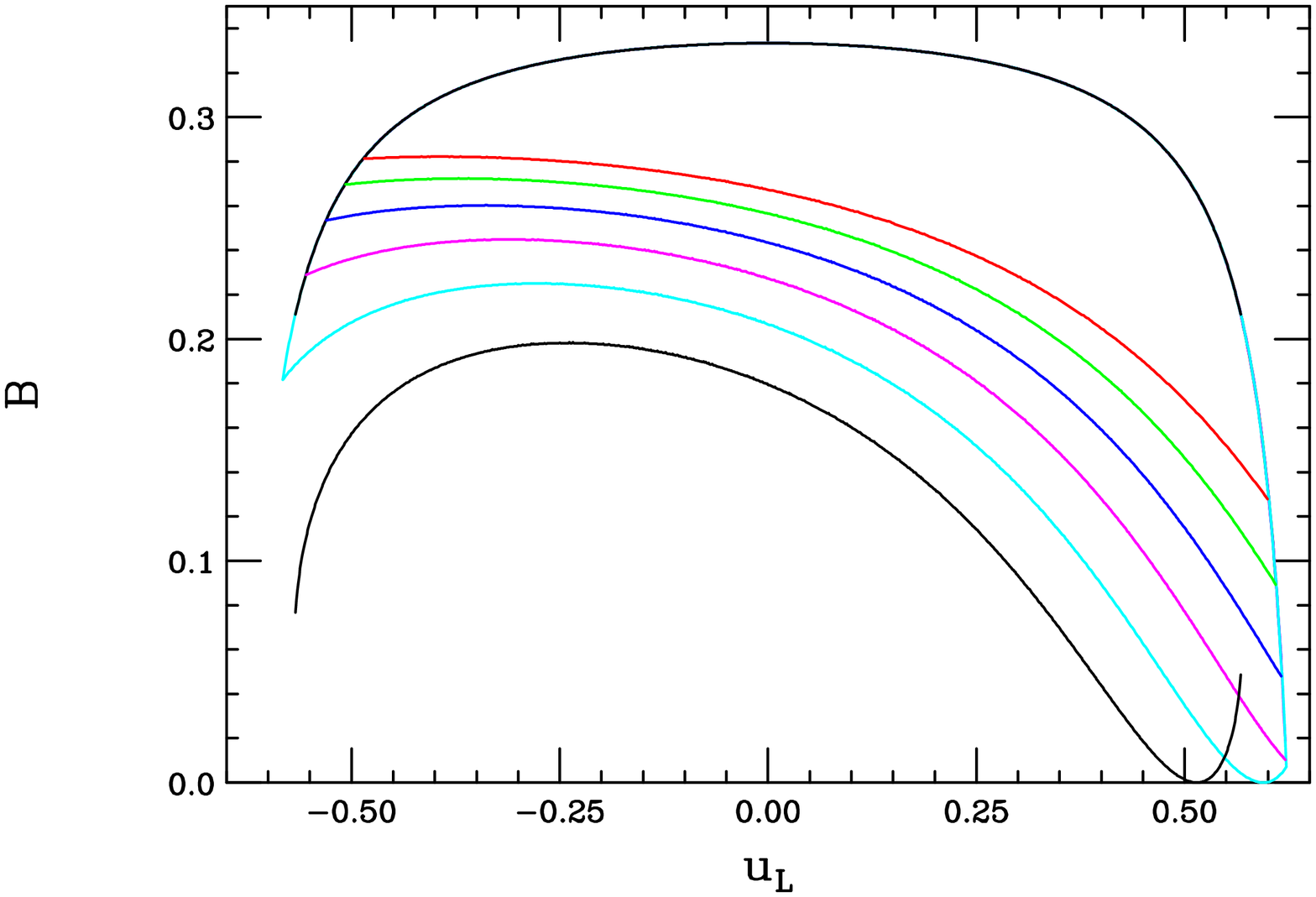}}
\vspace*{0.5cm}
\caption{(Top) Predicted ranges for the value of the $Z'$ width, $\Gamma$, arising from the parameter space along the color-coded arcs described above. 
(Bottom) The $b$-quark branching fraction of the $Z'$ for the corresponding range of coupling parameters.}
\label{width}
\end{figure}

To be consistent, we need to demonstrate that our use of the narrow width approximation is valid in the $Z'$ scenario. Essentially, it suffices to show that 
the $Z'$ total width, $\Gamma$, assuming decays to only SM particles, is always substantially smaller than the CDF dijet mass resolution, $\simeq 14.3$ 
GeV {\cite{Aaltonen:2011mk}}, for $M_{Z'}\sim 150$ GeV. Clearly, 
this condition will be most difficult to satisfy when the $Z'$ couples in a generation-independent manner to all 3 generations (as we have assumed here) 
instead of, \eg, only to the first generation which can then lead to significant flavor physics issues. Using the $Z'$ coupling parameter scans above, we can 
calculate the allowed regions for the predicted value of $\Gamma$; the results are shown in Fig.~\ref{width},
using the same color coding as before. Here 
we see that in the generation-independent coupling scenario, $\Gamma$ always remains in the range $0.5-5.6$ GeV, \ie, a set of values significantly below the 
CDF dijet mass resolution. Thus the $Z'$ will always appear to be 
quite narrow and, in particular, with $\Gamma/M_{Z'} \lsim 3.3\%$, validates our use of the narrow 
width approximation above. It is also of some interest to notice that the corresponding branching fractions for the decay $Z'\to b\bar b$, under the assumption 
of 3-generation coupling universality, are always found to lie in the approximate 
range $\sim 0-0.33$, which is consistent with the $Wjj$ data from CDF {\cite {blois,private}}. The coupling dependence of this branching fraction can be seen in 
detail in the lower panel of Fig.~\ref{width}. If lower values for this branching fraction are favored this would be an indication for couplings with 
$|u_L|>|d_L|$.

Since the above analysis restricts the allowed values of $u_L,d_L$ for the new $Z'$ boson (while $u_R,d_R$ play a lesser role and may in fact be zero) 
one would like to attempt to constrain these couplings further. Clearly a better determination of $\sigma_{WZ'}$ at the Tevatron and a measurement of 
$\sigma_{\gamma,Z+Z'}$ 
at both the Tevatron and LHC will be useful in this regard. However, it may be possible to obtain additional information from the $W+Z'$ kinematic 
distributions themselves. To this end, we return to our discussion of the $W+Z'$ differential cross section above. There, we saw that in the case of 
$u_L \neq d_L$ an additional term contributes to the cross section, \ie, the term denoted by $X$, which grows with increasing $\hat s$. If this term is absent, 
the differential distribution for $d\sigma/dM_{WZ'}$ will peak at low values of $M_{WZ'}$, not far above threshold and then fall rapidly. However, the 
presence of this term will push this peak in this distribution to significantly larger values of $M_{WZ'}$ and the corresponding 
fall off of this differential cross section 
will be far slower. Thus, in principle, a measurement of the $M_{WZ'}$ distribution could provide an additional useful handle on the $u_L,d_L$ coupling 
relationship which is independent of the values for $u_R,d_R$.

\begin{figure}[htbp]
\centerline{
\includegraphics[width=8.0cm,angle=90]{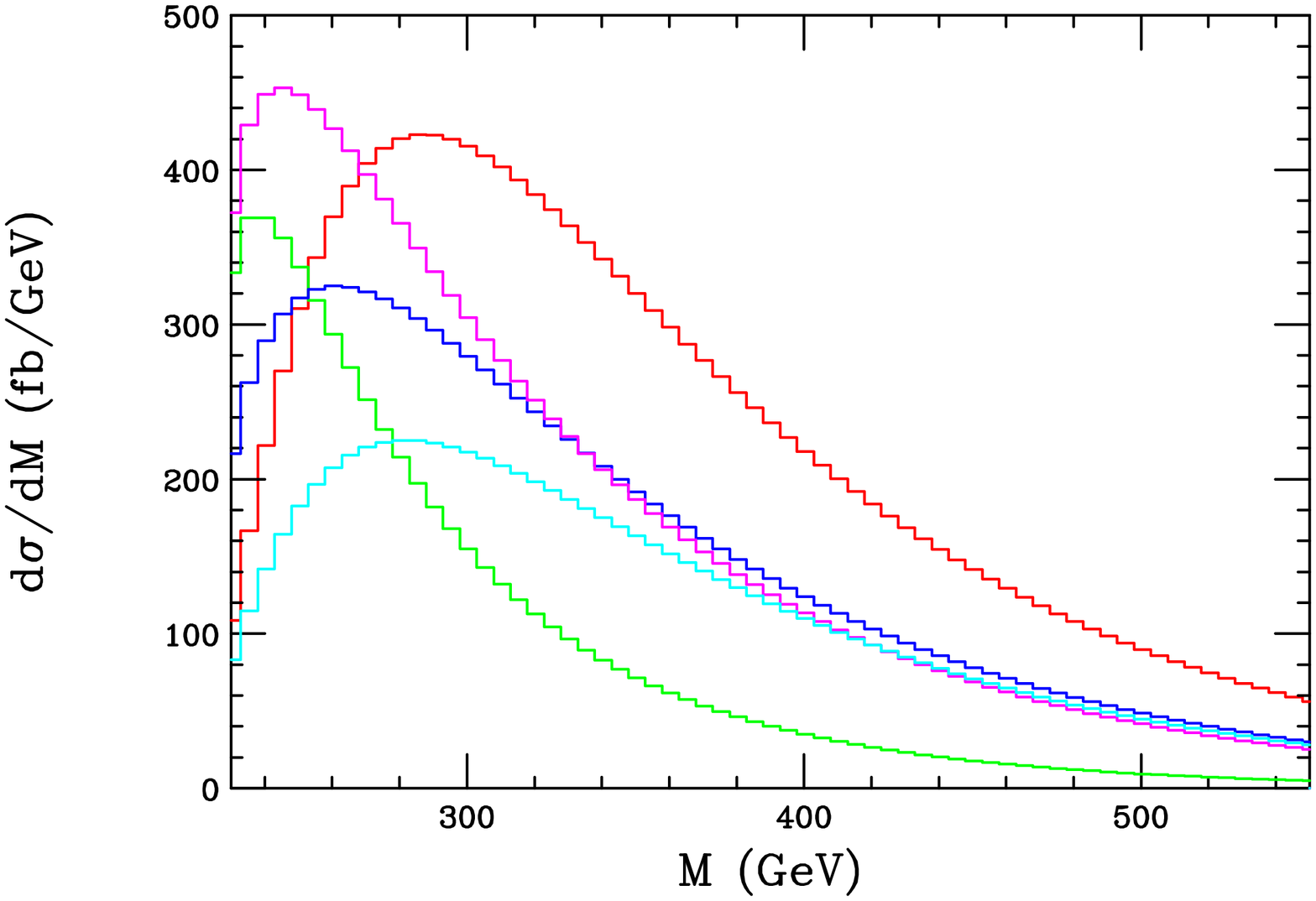}}
\vspace*{0.5cm}
\centerline{
\includegraphics[width=8.0cm,angle=90]{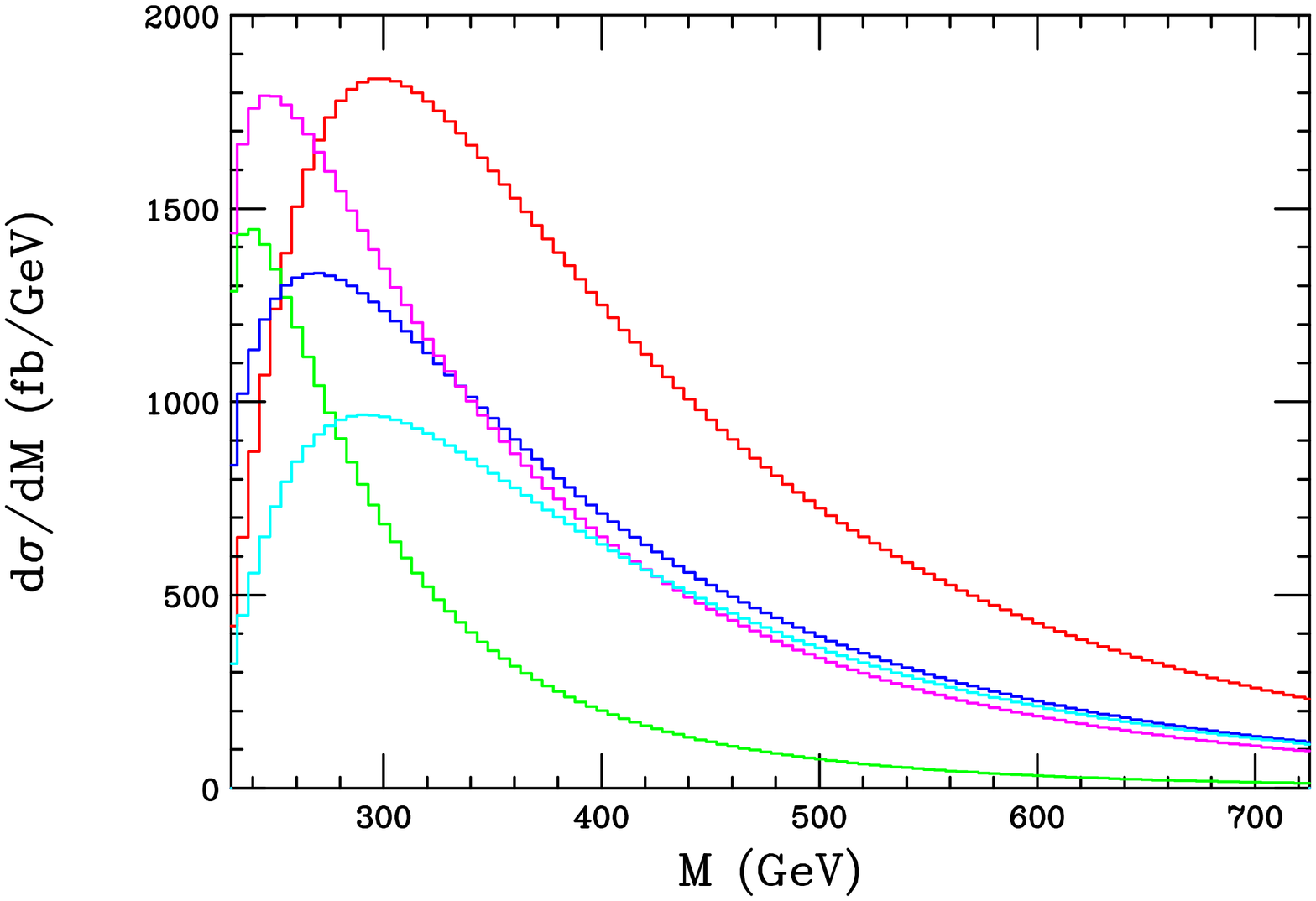}}
\vspace*{0.5cm}
\caption{$d\sigma/dM_{WZ'}$ distribution at the Tevatron(top) and LHC(bottom) for several different allowed values of $u_L,d_L$=(-0.5,0.5)(red),  
(-0.5,-0.5)(green), (0,0.7)(blue), (-0.2,-0.8)(magenta), (-0.2,0.5)(cyan), respectively.}
\label{generator}
\end{figure}

Since the `discovery' channel, $W^\pm Z'$, has the largeset cross section, a detailed study of this reaction can provide us a way to pin down the values of 
the $u_L,d_L$ couplings which will then further restrict $u_R,d_R$. 
In Fig.~\ref{generator} we show the $d\sigma/dM_{WZ'}$ distribution at the Tevatron and the LHC for several different representative values of $u_L,d_L$ 
lying within the allowed coupling ellipses shown in Fig.~\ref{ellipses}. In the top panel for $W+Z'$ production at the Tevatron, we see that this distribution 
is quite sensitive to the choice of these couplings. In particular, we see that when $u_L=d_L$ the peak in the distribution is at very low $M_{WZ'}$ values, not 
far from threshold, as expected.  However, the peak occurs at larger values of $M_{WZ'}$ when $u_L,d_L$ take on significantly different values. 
We especially note the strong differences between the  
cases of $u_L,d_L=(-0.5,0.5)$ and $u_L,d_L=(-0.5,-0.5)$. Fig.~\ref{generator} also shows the corresponding results for this distribution at the LHC which show 
similar coupling sensitivity since the shape of the distributions is quite similar to those found at the Tevatron. 

\begin{figure}[htbp]
\centerline{
\includegraphics[width=8.0cm,angle=90]{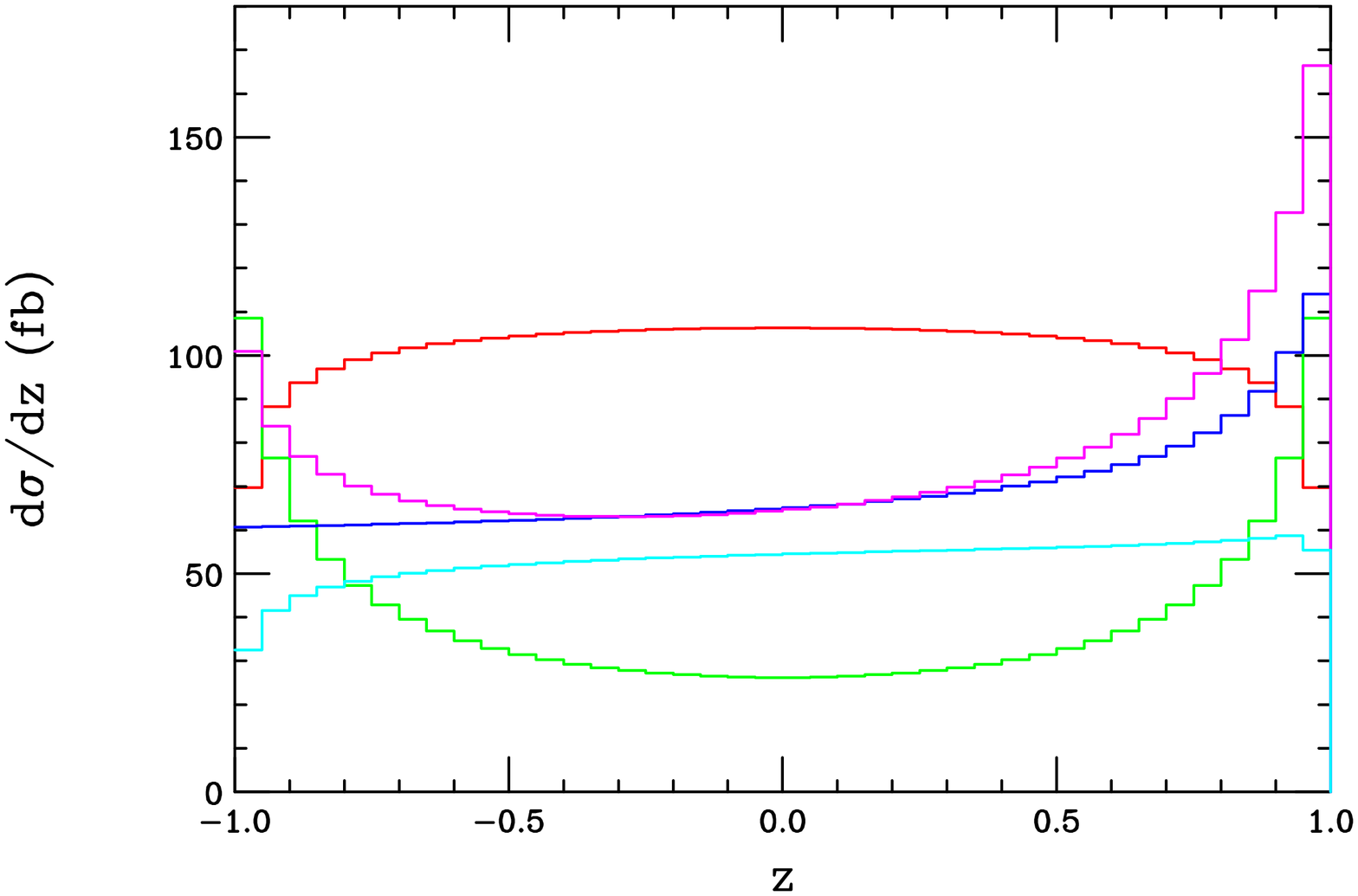}}
\vspace*{0.5cm}
\centerline{
\includegraphics[width=8.0cm,angle=90]{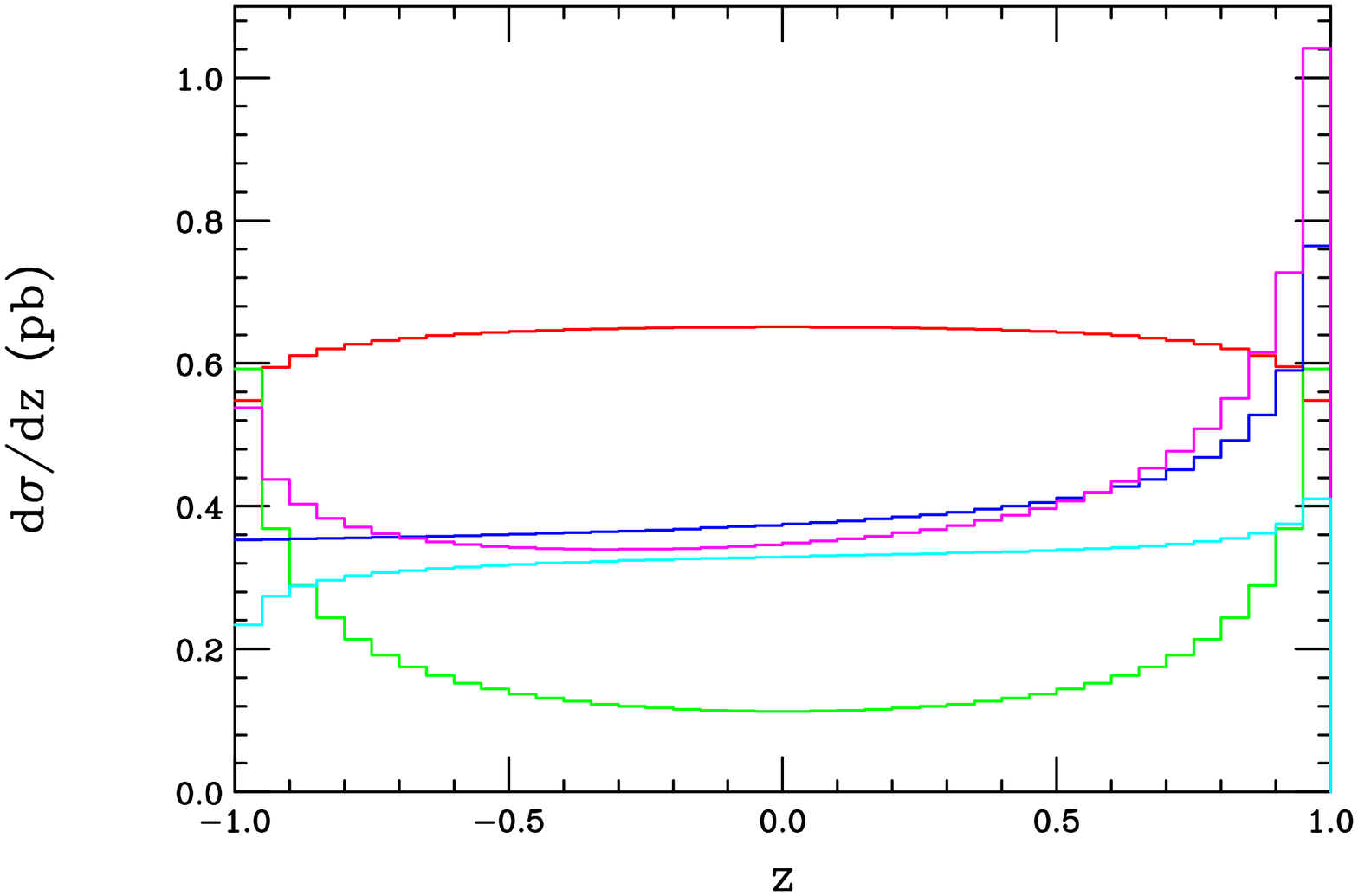}}
\vspace*{0.5cm}
\caption{$d\sigma/dz$ ($z=\cos \theta^*$, the CM scattering angle) distributions for the $W$ at the 
Tevatron(top) and LHC(bottom) for the same set of coupling choices as in the previous figure.}
\label{wzdist}
\end{figure}

Further information can be obtained by examining other kinematic distributions involving the $W^\pm$ or the dijet system. Fig.~\ref{wzdist} shows the angular 
distribution of the $W^\pm$ at both the Tevatron and the LHC. We notice several things: ($i$) The $d\sigma/dz$ distribution is very sensitive to the values of the 
$u_L,d_L$ couplings which can be traced back to the various terms in Eq.(2) above. First, we see that when $u_L=d_L$ the distribution is forward and backward 
peaked (due to the $u-$ and $t-$channel `poles') and is $z \to -z$ symmetric. In the other extreme, where the term proportional to $X$ in Eq.(2) dominates, we still have 
$z\to -z$ symmetry but the distribution is much flatter being proportional to $\sim \hat u \hat t$. In the intermediate cases where all terms are comparable, the 
$z\to -z$ symmetry is now lost and some forward and backward peaking is possible. However, the distributions are generally fairly flat for central values of $z$.   
($ii$) The angular distributions are quite different from what one would expect from scalar production. 
($iii$) As in the case of the $d\sigma/dM_{WZ'}$ spectrum, we note that the $W^\pm$ angular distributions look very similar at both colliders. This will remain 
true for the other distributions we display below and so we will only show the results for the LHC. 


\begin{figure}[htbp]
\centerline{
\includegraphics[width=8.0cm,angle=90]{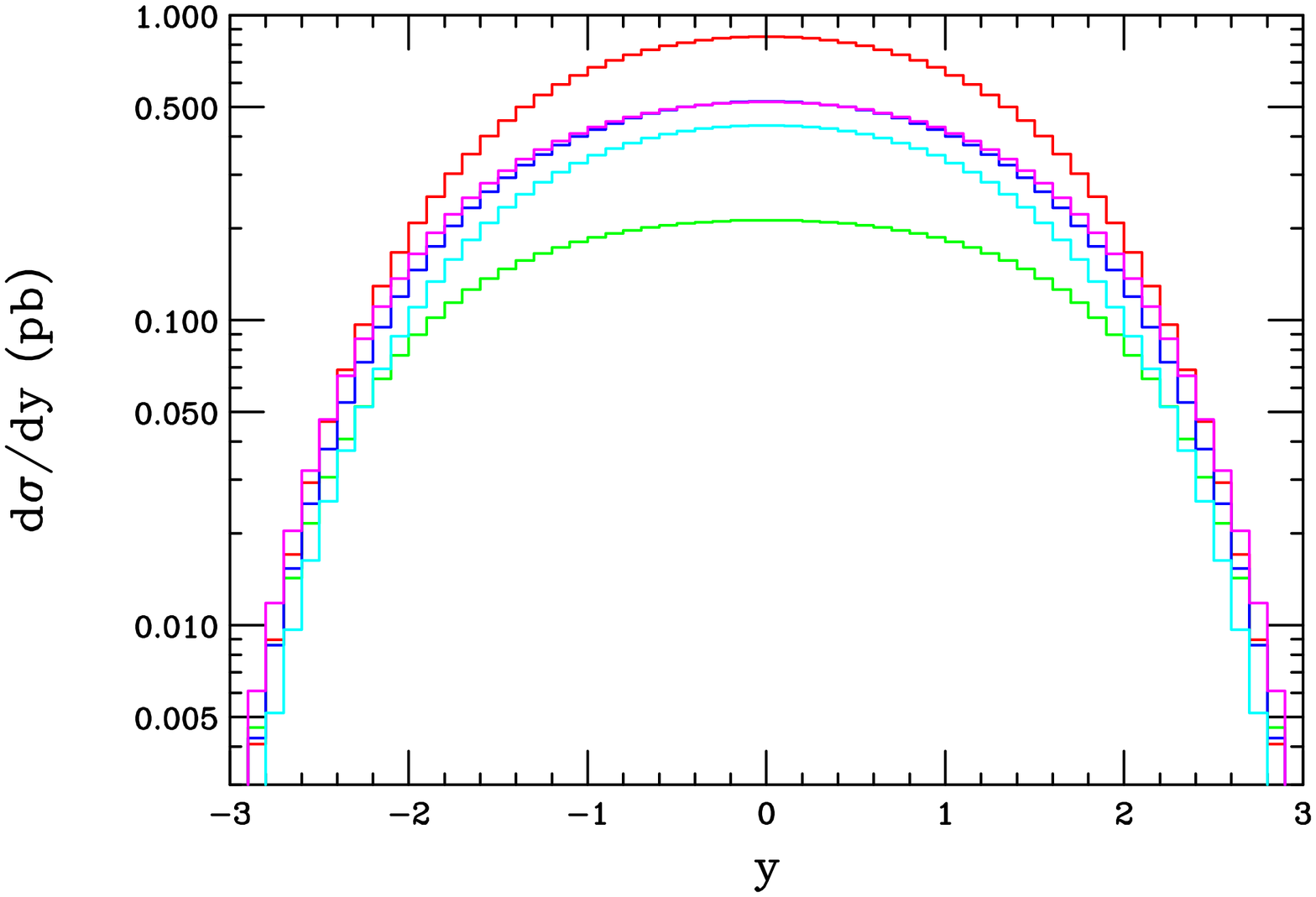}}
\vspace*{0.5cm}
\centerline{
\includegraphics[width=8.0cm,angle=90]{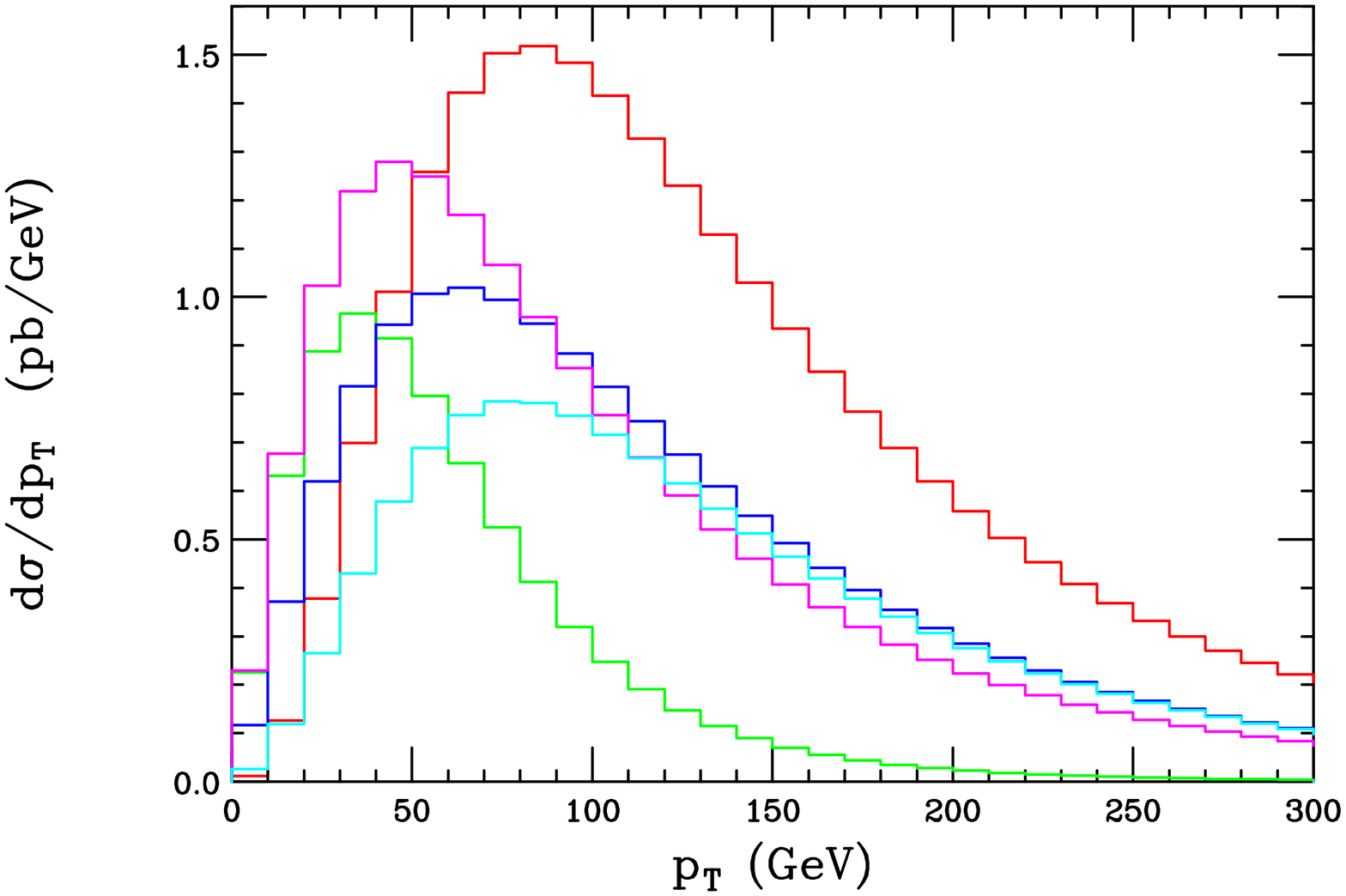}}
\vspace*{0.5cm}
\caption{Rapidity (top) and $p_T$(bottom) distributions for the $W^\pm$ at the LHC for the same set of coupling choices as in the previous figure.}
\label{wdist2}
\end{figure}

Figure~\ref{wdist2} shows both the $W^\pm$ rapidity ($y$) and $p_T$ distributions at the LHC for the same set of $u_L,d_L$ couplings as examined above. (As noted above, 
very similar results are obtained at the Tevatron.) The rapidity distribution shows only a relatively weak dependence on the couplings. However, it is easy to see 
that when $u_L=d_L$ it is quite flat in the central region, whereas, when $u_L$ and $d_L$ are very different it it much more peaked near $y=0$. On the otherhand, 
the $p_T$ spectrum of the $W^\pm$ is seen to be highly sensitive to the $u_L,d_L$ couplings as we might have expected based on the shapes of the $d\sigma/dM_{WZ'}$ 
and the $d\sigma/dz$ distributions discussed above. In particular we see that when $u_L \neq d_L$ the $W^\pm$ $p_T$ spectrum is somewhat harder, growing more so 
as the difference in couplings gets larger. Clearly information from this distribution will help in the determination of the left-handed quark couplings to the $Z'$.

\begin{figure}[htbp]
\centerline{
\includegraphics[width=8.0cm,angle=90]{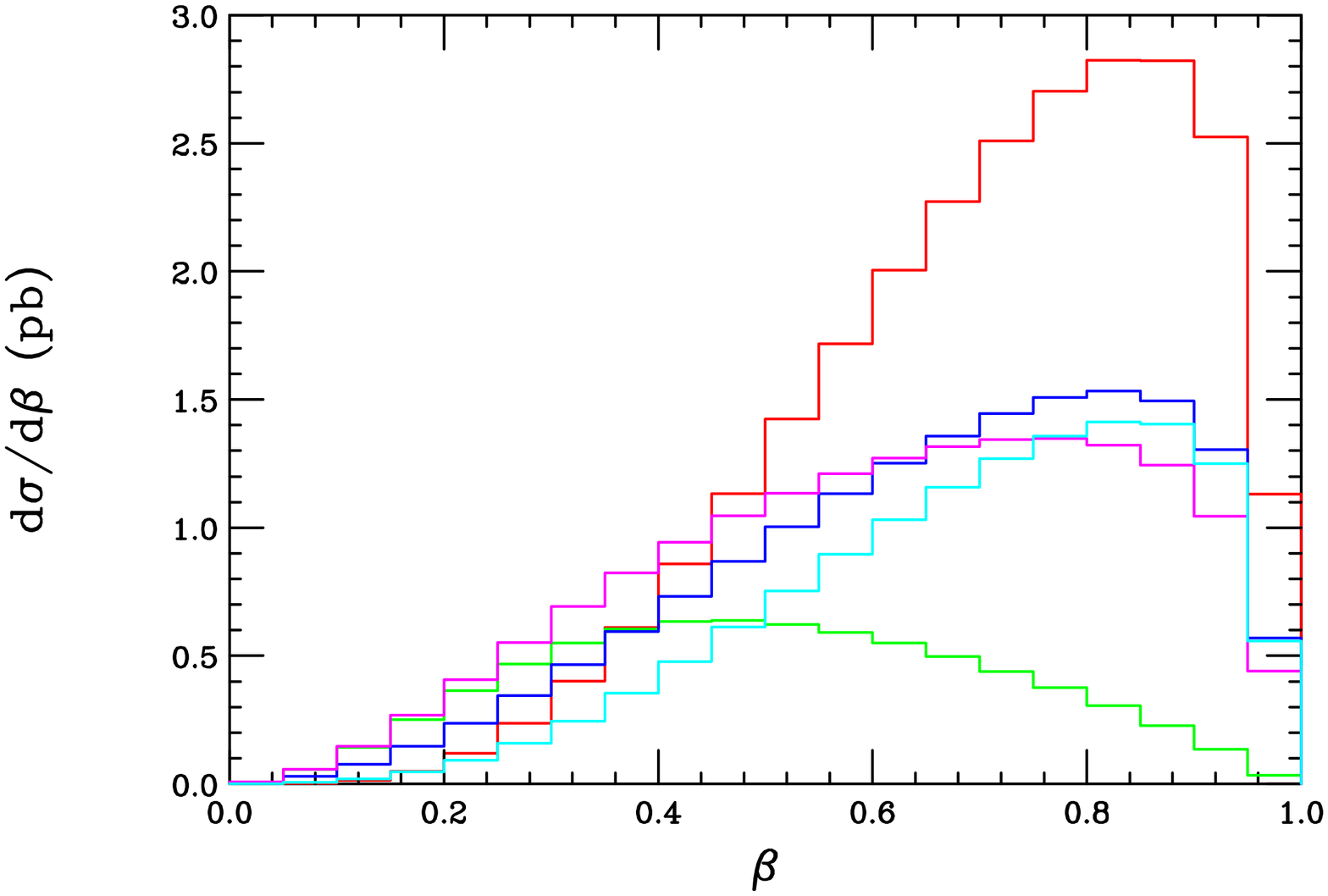}}
\vspace*{0.5cm}
\caption{Velocity ($\beta$) distribution for the $Z'$ at the LHC in the CM frame for the same set of coupling choices as in the previous figure.}
\label{wdist3}
\end{figure}

Figure.~\ref{wdist3} displays the velocity distribution of the $Z'$ in the CM frame; this is of particular importance in the determination of the boost required to go to 
the dijet CM frame in order to obtain the dijet angular distribution. A measurement of this quantity is necessary if one wants to verify the spin-1 nature of the $Z'$.  
Again, we see that this distribution is quite sensitive to the $u_L,d_L$ couplings and peaks at significantly larger values when $u_L \neq d_L$.

Lastly, we have also examined the possibility of observing $Z'$ bremsstrahlung in $q\bar q$ production in
$e^+e^-$ annihilation, \ie, $e^+e^-\to q\bar q Z'$.  We found that the rate for this
process is hopelessly small at LEPII energies and thus does not provide a constraint
on this scenario.

\section{Summary and Conclusions}

In summary, we have examined the hypothesis that a leptophobic $Z'$ boson accounts for the excess of events in the $Wjj$ channel as observed by CDF.
The quoted range for the production cross section places constraints on the left-handed couplings of the $Z'$ to the up- and down-quarks. Consistency with 
the lack of observation by D0 forces us to the lower end of this cross section range. Further consistency with the
non-observation of dijet resonances at $m_{jj}\sim 150$ GeV at UA2  constrains these couplings, and severely limits the possible values
of the $Z'$ right-handed couplings to the light quarks. Assuming that these couplings are generation independent, these 
results provide a relatively restrictive allowed region for the four hadronic couplings
of the $Z'$.  

These allowed coupling regions translate into well-determined 
rates for the associated production of $Z/\gamma+Z'$ at the Tevatron and LHC, as well as for $W+Z'$ at the LHC, apart from NLO corrections. 
The $Wjj$ rate at the LHC is large and this channel
should be observed soon once the SM backgrounds are under control. The rates for $Z/\gamma+Z'$ associated production are smaller, and these processes
should not yet have been observed at the Tevatron given the expected SM backgrounds.  Once detected, these processes will provide valuable information
on the $Z'$ boson couplings. Further information on the $u_L-d_L$ coupling relationship was shown to also be obtainable from measurements of the 
$d\sigma/dM_{WZ'}$ as well as other kinematic distributions at both the Tevatron and the LHC.

Even with four free coupling parameters, this scenario is predictive, even more so once the $W+Z'$ cross section is better determined  at the Tevatron, and can be 
further tested at both the Tevatron itself as well as at the LHC in the near future.  In particular, the LHC should confirm (or not) this scenario soon.
\newpage

\noindent{\Large\bf Acknowledgments}\\

The authors would like to thank Viviana Cavaliere for discussions and for answering so many of our questions about the CDF $Wjj$ analysis. The authors 
would also like to thank Yang Bai for discussions about the $Wjj$ distributions.

%
\def\IJMP #1 #2 #3 {Int. J. Mod. Phys. A {\bf#1},\ #2 (#3)}
\def\MPL #1 #2 #3 {Mod. Phys. Lett. A {\bf#1},\ #2 (#3)}
\def\NPB #1 #2 #3 {Nucl. Phys. {\bf#1},\ #2 (#3)}
\def\PLBold #1 #2 #3 {Phys. Lett. {\bf#1},\ #2 (#3)}
\def\PLB #1 #2 #3 {Phys. Lett. B {\bf#1},\ #2 (#3)}
\def\PR #1 #2 #3 {Phys. Rep. {\bf#1},\ #2 (#3)}
\def\PRD #1 #2 #3 {Phys. Rev. D {\bf#1},\ #2 (#3)}
\def\PRL #1 #2 #3 {Phys. Rev. Lett. {\bf#1},\ #2 (#3)}
\def\PTT #1 #2 #3 {Prog. Theor. Phys. {\bf#1},\ #2 (#3)}
\def\RMP #1 #2 #3 {Rev. Mod. Phys. {\bf#1},\ #2 (#3)}
\def\ZPC #1 #2 #3 {Z. Phys. C {\bf#1},\ #2 (#3)}

\end{document}